\documentclass[3p]{elsarticle}

\usepackage{pifont, geometry, fleqn, graphicx}
\usepackage[colorlinks,bookmarksopen,bookmarksnumbered,citecolor=red,urlcolor=red]{hyperref}
\RequirePackage[labelfont={rm,md,up},format=hang]{subfig}
\usepackage{colortbl}
\usepackage{xcolor}
\usepackage{hhline}
\usepackage[USenglish]{babel}
\usepackage{amsmath}
\usepackage{amsfonts}
\usepackage{amssymb}
\usepackage{mathrsfs}

\newcommand\BibTeX{{\rmfamily B\kern-.05em \textsc{i\kern-.025em b}\kern-.08em
T\kern-.1667em\lower.7ex\hbox{E}\kern-.125emX}}

\makeatletter
\def\ps@pprintTitle{%
 \let\@oddhead\@empty
 \let\@evenhead\@empty
 \def\@oddfoot{}%
 \let\@evenfoot\@oddfoot}
\makeatother

\begin{document}

\begin{frontmatter}

\title{De-homogenization of optimal multi-scale 3D topologies}

\author[dep1]{Jeroen P. Groen\corref{cor1}\fnref{ft1}}
\author[dep2]{Florian C. Stutz} 
\author[dep1]{Niels Aage}
\author[dep2]{Jakob A. B{\ae}rentzen}
\author[dep1]{Ole Sigmund}

\fntext[ft1]{E-mail: jergro@mek.dtu.dk}
\cortext[cor1]{Correspondence to: J. P. Groen, Department of Mechanical Engineering, Solid Mechanics, Technical University of Denmark, Nils Koppels All\'{e}, Building 404, 2800 Kgs. Lyngby, Denmark}

\address[dep1]{Department of Mechanical Engineering, Solid Mechanics, Technical University of Denmark}
\address[dep2]{Department of Applied Mathematics and Computer Science, Technical University of Denmark}

\begin{abstract}
\begin{sloppypar}
This paper presents a highly efficient method to obtain high-resolution, near-optimal 3D topologies optimized for minimum compliance on a standard PC. Using an implicit geometry description we derive a single-scale interpretation of optimal multi-scale designs on a very fine mesh (de-homogenization). By performing  homogenization-based topology optimization, optimal multi-scale designs are obtained on a relatively coarse mesh resulting in a low computational cost. As microstructure parameterization we use orthogonal rank-3 microstructures, which are known to be optimal for a single loading case. Furthermore, a method to get explicit control of the minimum feature size and complexity of the final shapes will be discussed. Numerical examples show excellent performance of these fine-scale designs resulting in objective values similar to the homogenization-based designs. Comparisons with well-established density-based topology optimization methods show a reduction in computational cost of 3 orders of magnitude, paving the way for giga-scale designs on a standard PC.
\end{sloppypar}
\end{abstract}


\end{frontmatter}

\section{Introduction}
Topology optimization is an advanced design tool with the power to provide engineers with novel insights about optimal design. Nowadays, availability of high performance computing resources allows for the application of topology optimization on realistic design problems using different types of physics. Examples include, optimization of heat sinks using natural convection~\citep{Bib:AlexandersenHeat2016}, optimizing fluid flow systems by modeling turbulence using Reynolds-averaged Navier-Stokes~\citep{Bib:DilgenDilgen2018}, and acoustic horn optimization~\citep{Bib:UmeaHornOpt2016}.
However, the most studied (and arguably the most simple) type of optimization problem is compliance minimization ($i.e.$ stiffness maximization) of linear elastic structures subject to one or more load-cases and an upper bound on the material useage. Recently,~\citet{Bib:AageNature2017} extended the state-of-the-art of solution methods for these type of problems by optimizing an airplane wing using more than 1 billion design variables. To do so, 8000 cores were employed on a high performance computer system up to 5 days. Hence, a significant reduction in computational cost is still required if large-scale topology optimization is to be adapted interactively in engineering practice.

Theoretically, it is known that the optimal shape for a compliance minimization problem contains periodic details on several
length-scales~\citep{Bib:KohnStrang1986,Bib:AllaireAubry1999}. Instead of modeling these microscopic details on an extremely fine mesh, the problem can be decomposed into a multi-scale problem. The theory of homogenization can be used to calculate the effective macroscopic properties of these complex but periodic microstructures~\citep{Bib:Bensoussan1978}. A class of microstructures that is able to reach the bounds on maximum strain energy are the so-called rank-N laminates~\citep{Bib:Lurie1984,Bib:Norris1985,Bib:Milton86,Bib:FrancfortMurat86}. These composites, which contain several length-scales, have the nice property that their corresponding elasticity tensors can be analytically derived. By using homogenization-based topology optimization the design space can be relaxed to allow for these rank-N laminates~\citep{Bib:Bendsoe1989}, such that optimal designs can be obtained at a much lower computational cost compared to density-based topology optimization. The theory for performing homogenization-based topology optimization using optimal microstructures is well-established and described in detail in the books by~\citet{Bib:AllaireBook,Bib:TopOptBook}. Nevertheless, the interest in this method has faded in the last decade, which can be explained by the fact that the optimal designs are on several length-scales which prohibits their manufacturing. 

To avoid the use of multi-scale microstructures, it is possible to use a database of near optimal single-scale microstructures whose properties can be obtained using numerical homogenization~\citep{Bib:BendsoeKikuchi}. Furthermore, it is possible to perform the optimization procedure in a hierarchical sense, which means that both the material distribution and single-scale microstructures are optimized iteratively by making use of inverse homogenization~\citep{Bib:SigmundInverse1994,Bib:RodriguesHierarchical}. Unfortunately, such an approach comes at large computational cost~\citep{Bib:CoelhoParallelHierarchical}. To reduce this cost~\citet{Bib:LiuPAMP2008} and~\citet{Bib:Sivapuram} restricted the number of different microstructures in the design domain; however, this results in designs far away from the true optimum. Furthermore, the optimized unit-cell designs do not take into account connectivity and load transfer between adjacent microstructures due to the separation of scales. In fact, it is possible to reconstruct near optimal single-scale microstructures based on optimal rank-N laminates~\citep{Bib:TraffSigmundGroen2019}. This approach results in simpler microstructures than the ones obtained using inverse homogenization, at a negligible computational cost. Furthermore,~\citet{Bib:PantzTrabelsi} showed that a multi-scale design can be interpreted on a single-scale using an implicit geometry description. Recent works on this so-called "de-homogenization" approach have shown that high-resolution near-optimal 2D structures subject to a single loading case can be obtained at a fraction of the cost of density-based topology optimization~\citep{Bib:GroenSigmund2017,Bib:AllaireDondersPantz2D,Bib:GroenWuSigmund2019}. 

This article presents a natural extension of the de-homogenization approach for elasticity problems in 3D subject to a single loading case. Optimal rank-3 microstructures are used for the homogenization-based topology optimization, and the multi-scale designs are subsequently de-homogenized on fine meshes containing more than $200$ million voxels. Using this approach large-scale designs can be efficiently obtained on a modern PC, without the need for a high performance computing system, allowing topology optimization to become a more integrated part of the design process. Finally, it should be noted that simultaneous to this study different research groups have been working on related type of methods.~\citet{Bib:Perle3D} present an approach in which sub-optimal open-walled microstructures are de-homogenized on a fine grid using a slightly different approach.~\citet{Bib:Wu3DLattice2019} present an approach to reconstruct a conformal lattice design from a homogenization-based design without creating a global parameterization. Contrary to above studies we use optimal rank-3 microstructures, since these microstructures have much better load carrying capabilities than sub-optimal truss-like microstructures~\citep{Bib:SigmundMichell}.

The article is organized as follows: the rank-3 parameterization and the methodology to do homogenization-based topology optimization is presented in Section~\ref{Sec:Hbased}. The de-homogenization procedure and a method to control the shape of the final designs is presented in Section~\ref{Sec:DeHomo}. A large number of numerical examples to demonstrate the performance of the presented approach as well as extensive comparisons with density-based topology optimization are presented in Section~\ref{Sec:NumExp}. Finally, the most important conclusions of this study will be discussed in Section~\ref{Sec:Conclusion}.
\section{Homogenization-based topology optimization}
\label{Sec:Hbased}
 
\subsection{The elasticity tensor of a rank-3 laminate}
\begin{figure}[h!]
\centering
\subfloat[Rank-1 laminate in local frame.]{\includegraphics[width=0.328\textwidth]{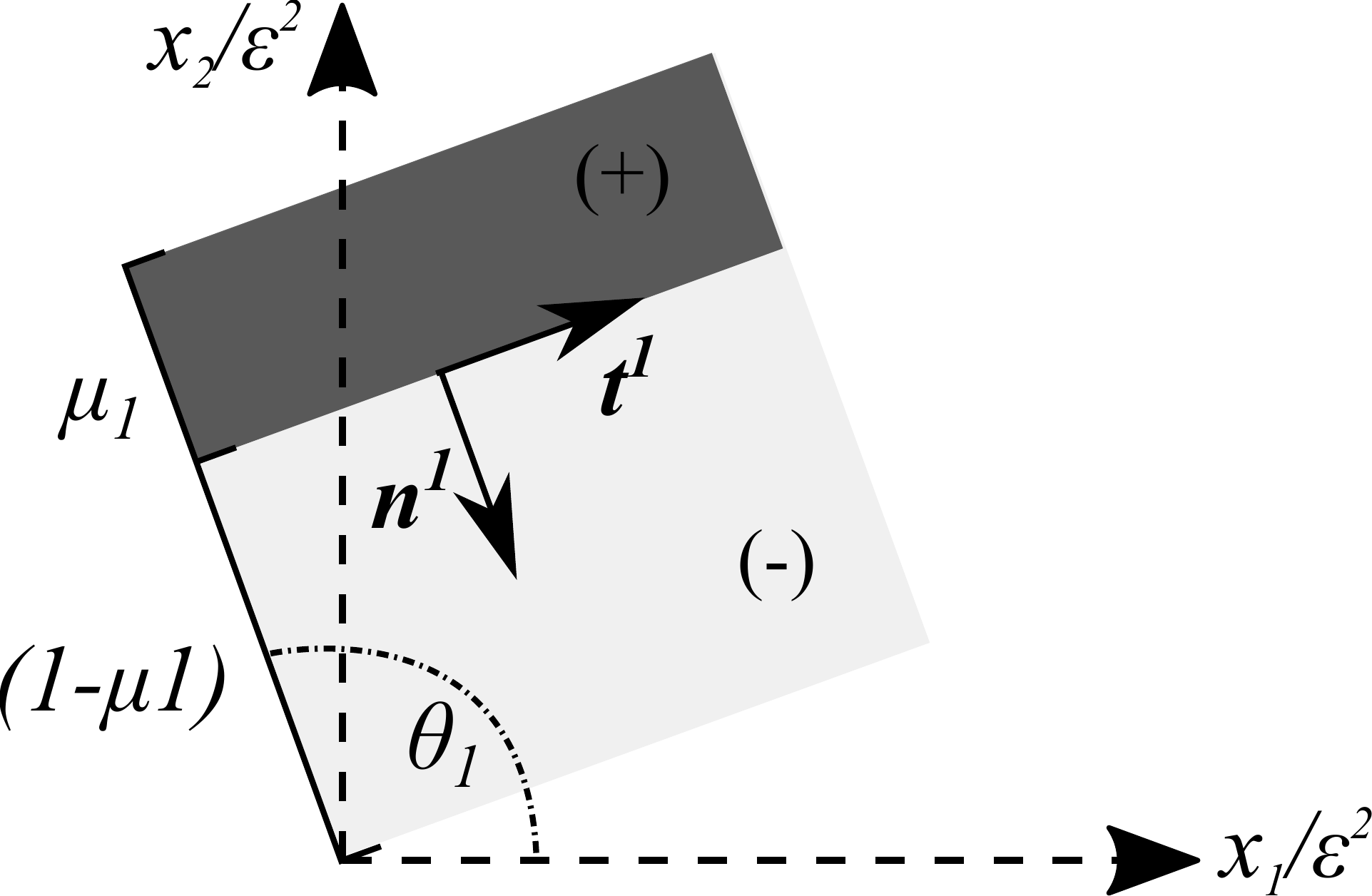}} \hfill
\subfloat[Rank-2 laminate in local frame.]{\includegraphics[width=0.328\textwidth]{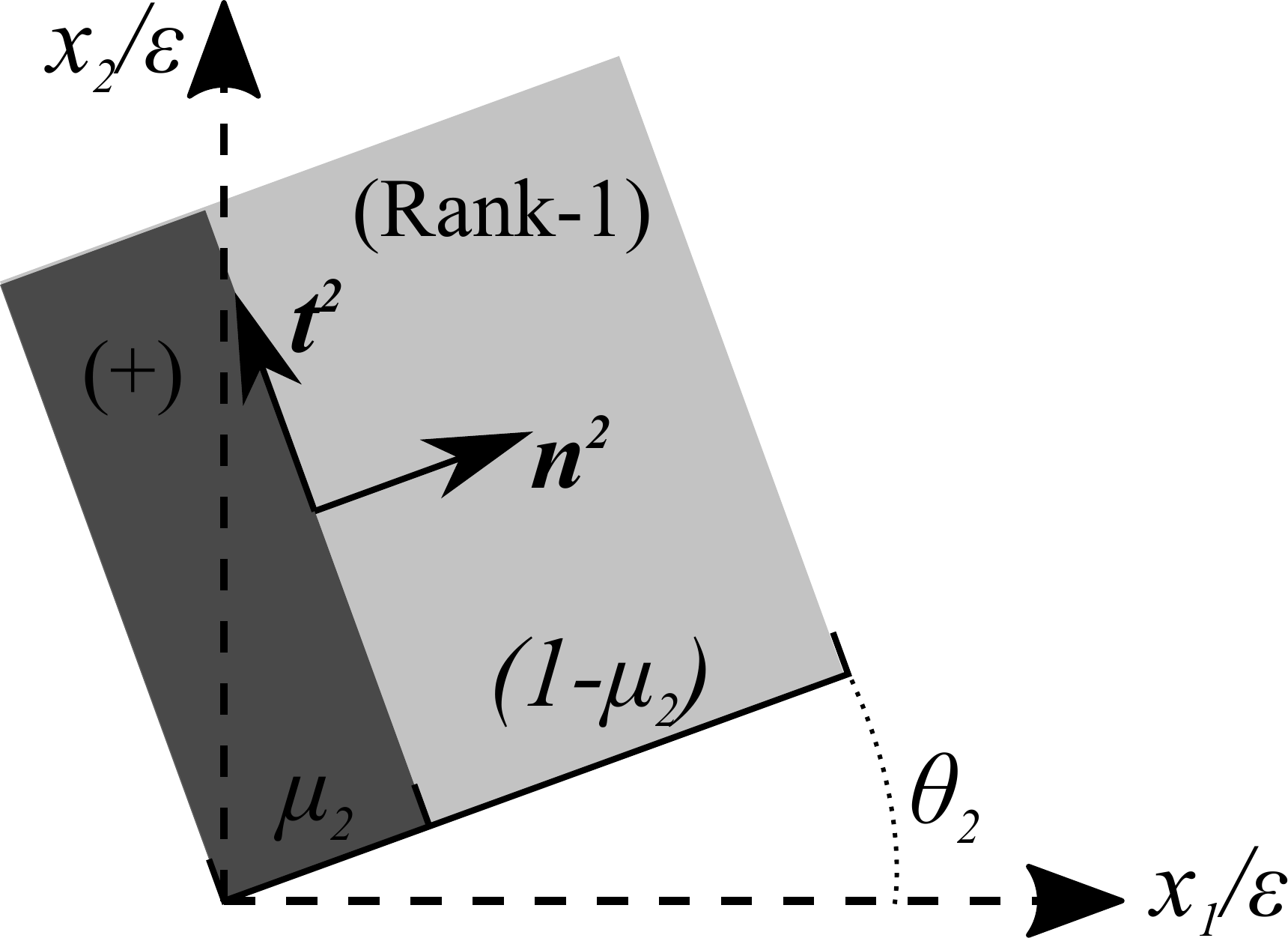}} \hfill
\subfloat[Periodic rank-2 laminate.]{\includegraphics[width=0.238\textwidth]{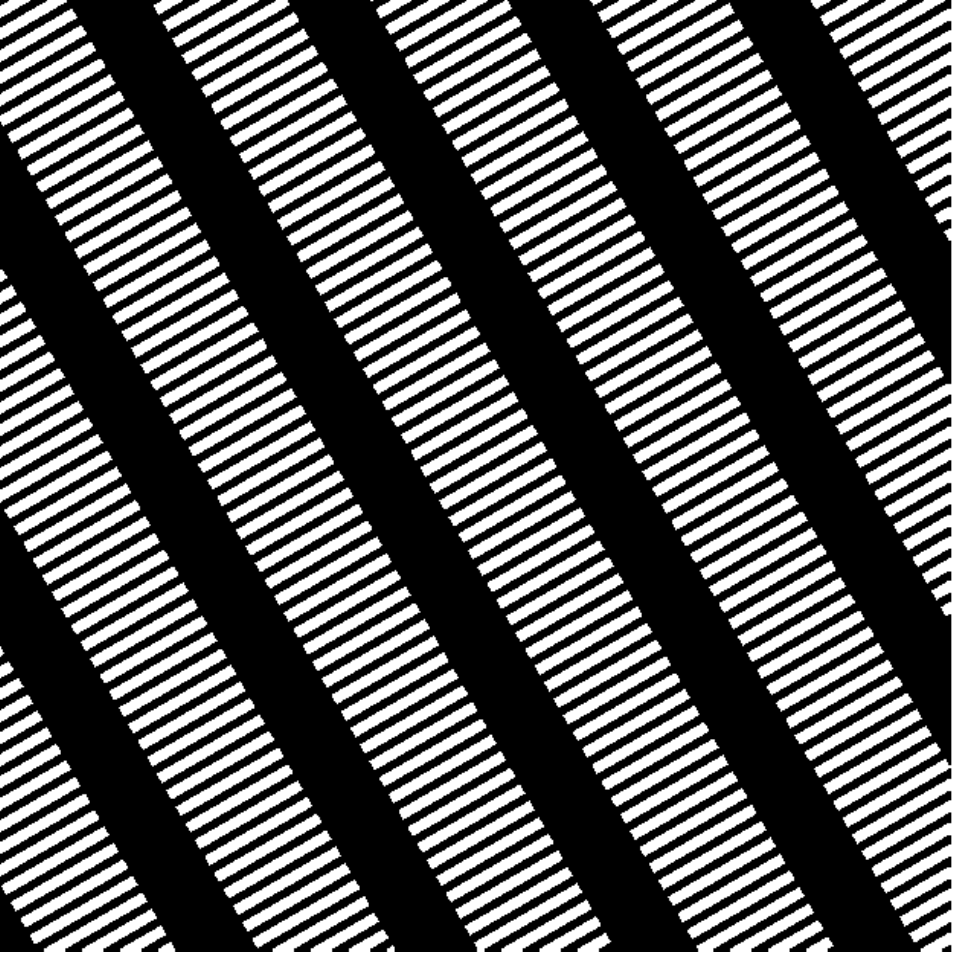}}
\caption{Visualization of how a rank-2 microstructure is constructed from a stiff isotropic material (+) and a weak/compliant isotropic material (-). Please note the different length-scales.}
\label{Fig:Rank3.1}
\end{figure}

It is well-known that the optimal design for compliance minimization problems subject to a single loading case can be described by macroscopically varying orthogonal rank-3 laminates~\citep{Bib:FrancfortMurat86,Bib:Avellaneda87,Bib:GibianskyCherkaev1987}. These multi-scale microstructures consist of two different materials, a stiff isotropic material (+) and a weak/compliant isotropic material (-) mimicking void, using Young's moduli $E^{+}$ and $E^{-}$, respectively, and both with identical Poisson's ratio $\nu$. To visualize such a laminate, we consider the orthogonal rank-2 laminate shown in Figure~\ref{Fig:Rank3.1}, which is optimal for a planar problem subject to a single load case. The first layering of this microstructure is a periodic rank-1 laminate, shown in Figure~\ref{Fig:Rank3.1} (a). The orientation of the rank-1 laminate is described by the layer normal $\boldsymbol{n}^{1}$ and layer tangent $\boldsymbol{t}^{1}$. Furthermore, the relative layer width of the stiff material is described by parameter $\mu_{1} \in [0,1]$, hence the layer width of the compliant material is $(1-\mu_{1})$. The elasticity tensor $\boldsymbol{E}^{R1}$ of the rank-1 laminate can be analytically calculated using the theory of homogenization, by assuming a periodic microstructure and perfect bonding between the two material phases, see e.g.~\citep{Bib:AllaireBook,Bib:TopOptBook}. 

The rank-1 microstructure is shown in relation to the global frame of reference $(x_{1},x_{2})$; however, please note that the depicted length-scale is $\boldsymbol{x}/\varepsilon^{2}$ with $\varepsilon \rightarrow 0$. This means that the microstructure can be assumed  uniform at the length-scale $\boldsymbol{x}/\varepsilon$. The rank-2 microstructure is then constructed on this length-scale, ($i.e.$ $\boldsymbol{x}/\varepsilon$) by combining the stiff material phase and the rank-1 microstructure as is shown in Figure~\ref{Fig:Rank3.1} (b). By setting $\boldsymbol{n}^{2} = \boldsymbol{t}^{1}$ and $\boldsymbol{t}^{2} = \boldsymbol{n}^{1}$ orthogonality of the layers is enforced. In a similar fashion as before the elasticity tensor  $\boldsymbol{E}^{R2}$ of the rank-2 laminate can be analytically derived as a function of orientation angle $\theta_{2}$ and relative layer widths $\mu_{1}$ and $\mu_{2}$.

The procedure for constructing a rank-3 laminate in 3D is exactly the same; however, now each layer $i$ has two tangents $\boldsymbol{t}^{i,1}$ and $\boldsymbol{t}^{i,2}$ to describe the plane spanned by the stiff material. The elasticity tensor for a rank-3 laminate $\boldsymbol{E}^{R3}$ can be described by,
\begin{equation}\label{Eq:Rank3.1}
\begin{aligned}
\boldsymbol{E}^{R3}&= \boldsymbol{E}^{+}-(1-\mu_{1})(1-\mu_{2})(1-\mu_{3}) 
 \Big((\boldsymbol{E}^{+}-\boldsymbol{E}^{-})^{-1} - \\
&(1+\nu)(1-2\nu)\frac{\mu_{1} \boldsymbol{\Lambda}(\boldsymbol{n}^{1},\boldsymbol{t}^{1,1},\boldsymbol{t}^{1,2}) + (1-\mu_{1}) (\mu_{2} \boldsymbol{\Lambda}(\boldsymbol{n}^{2},\boldsymbol{t}^{2,1},\boldsymbol{t}^{2,2}) + (1-\mu_{2})\mu_{3} \boldsymbol{\Lambda}(\boldsymbol{n}^{3},\boldsymbol{t}^{3,1},\boldsymbol{t}^{3,2}))} {E^{+}} \Big)^{-1}.
\end{aligned}
\end{equation}
The elasticity tensor is thus described by the layer normals $\boldsymbol{n}^{i}$ and tangents $\boldsymbol{t}^{i,1},\boldsymbol{t}^{i,2}$ as well as the relative layer widths $\mu_{i}$, for $i = 1,2,3$. The influence of the orientation of each layer is captured by the fourth-order tensor $\boldsymbol{\Lambda}$,
\begin{equation}\label{Eq:Rank3.2}
\begin{aligned}
\boldsymbol{\Lambda}(\boldsymbol{n}^{i},&\boldsymbol{t}^{i,1},\boldsymbol{t}^{i,2})  = \frac{1}{(1-\nu)} \boldsymbol{n}^{i} \otimes \boldsymbol{n}^{i} \otimes \boldsymbol{n}^{i}  \otimes \boldsymbol{n}^{i} + \frac{1}{2(1-2\nu)}   \\
& \Big( \boldsymbol{t}^{i,1} \otimes \boldsymbol{n}^{i} \otimes \boldsymbol{t}^{i,1}  \otimes \boldsymbol{n}^{i} + \boldsymbol{n}^{i} \otimes \boldsymbol{t}^{i,1} \otimes \boldsymbol{t}^{i,1}  \otimes \boldsymbol{n}^{i} + \boldsymbol{t}^{i,1} \otimes \boldsymbol{n}^{i} \otimes \boldsymbol{n}^{i}  \otimes \boldsymbol{t}^{i,1}  +  \boldsymbol{n}^{i} \otimes \boldsymbol{t}^{i,1} \otimes \boldsymbol{n}^{i}  \otimes \boldsymbol{t}^{i,1}  +\\
& \boldsymbol{t}^{i,2} \otimes \boldsymbol{n}^{i} \otimes \boldsymbol{t}^{i,2}  \otimes \boldsymbol{n}^{i} + \boldsymbol{n}^{i} \otimes \boldsymbol{t}^{i,2} \otimes \boldsymbol{t}^{i,2}  \otimes \boldsymbol{n}^{i} + \boldsymbol{t}^{i,2} \otimes \boldsymbol{n}^{i} \otimes \boldsymbol{n}^{i}  \otimes \boldsymbol{t}^{i,2}  +  \boldsymbol{n}^{i} \otimes \boldsymbol{t}^{i,2} \otimes \boldsymbol{n}^{i}  \otimes \boldsymbol{t}^{i,2}
\Big),
\end{aligned}
\end{equation}
where $\otimes$ indicates the dyadic product. The normal and tangent vectors of the three layers are linked since we use an orthogonal rank-3 microstructure. It is well-known that three Euler angles are required to represent an orthogonal frame in 3D. Hence, we use angles $\theta_{1},\theta_{2}$ and $\theta_{3}$ to define the different layer normals and tangents,
\begin{equation}\label{Eq:Rank3.3}
\begin{aligned}
\boldsymbol{n}^{1} & = \boldsymbol{t}^{2,1} = \boldsymbol{t}^{3,1} = \begin{Bmatrix} \text{cos}(\theta_{1})\text{cos}(\theta_{3})\text{sin}( \theta_{2}) + \text{sin}(\theta_{1})\text{sin}(\theta_{3}) \\
\text{cos}(\theta_{1})\text{sin}(\theta_{2})\text{sin}(\theta_{3})-\text{cos}(\theta_{3})\text{sin}(\theta_{1}) \\
\text{cos}(\theta_{1})\text{cos}(\theta_{2})\end{Bmatrix}, \\
\boldsymbol{n}^{2} &= \boldsymbol{t}^{1,1} = \boldsymbol{t}^{3,2} = 
\begin{Bmatrix}
\text{cos}(\theta_{3})\text{sin}(\theta_{1})\text{sin}(\theta_{2})-\text{cos}(\theta_{1})\text{sin}(\theta_{3}) \\
\text{sin}(\theta_{1})\text{sin}(\theta_{2})\text{sin}(\theta_{3})+ \text{cos}(\theta_{1})\text{cos}(\theta_{3}) \\
\text{sin}(\theta_{1})\text{cos}(\theta_{2})
\end{Bmatrix}, \\
\boldsymbol{n}^{3} &= \boldsymbol{t}^{1,2} = \boldsymbol{t}^{2,2} = 
\begin{Bmatrix}
\text{cos}(\theta_{2})\text{cos}(\theta_{3})\\
\text{cos}(\theta_{2})\text{sin}(\theta_{3})\\
-\text{sin}(\theta_{2})
\end{Bmatrix}.
\end{aligned}
\end{equation}
This means that the elasticity tensor of a rank-3 laminate can be described by six parameters $\mu_{1},\mu_{2},\mu_{3},\theta_{1},\theta_{2}$ and $\theta_{3}$. And it is possible to find the derivatives of $\boldsymbol{E}^{R3}$ w.r.t to these variables to perform gradient-based optimization. These expressions for the gradients are long but not necessarily difficult to derive; furthermore, the material volume fraction $\rho$ of a rank-3 microstructure is defined as,
\begin{equation}\label{Eq:Rank3.4}
\rho = \mu_{1} + \mu_{2} + \mu_{3} - \mu_{1}\mu_{2} - \mu_{1}\mu_{3} - \mu_{2}\mu_{3} + \mu_{1}\mu_{2}\mu_{3}.
\end{equation}

\subsection{Regularization of layer widths and avoiding thin features}
The homogenization-based topology optimization problem will be solved on a finite element mesh discretized using tri-linear finite elements, which each hold a uniform microstructure. As is shown by~\citet{Bib:DiazSigmund1995} a checkerboard-like pattern analyzed using linear finite elements can be stiffer than an optimal microstructure with the same average density. To avoid these artificially stiff patterns, a classical density filter is used to obtain filtered relative layer widths $\tilde{\mu}_{i}$ from $\mu_{i}$~\citep{Bib:DensityBourdin,Bib:DensityBruns}. As filter radius we use $R=1.5~h^{c}$, with $h^{c}$ the length of an element.

\begin{figure}[ht!]
\includegraphics[width=1.0\textwidth]{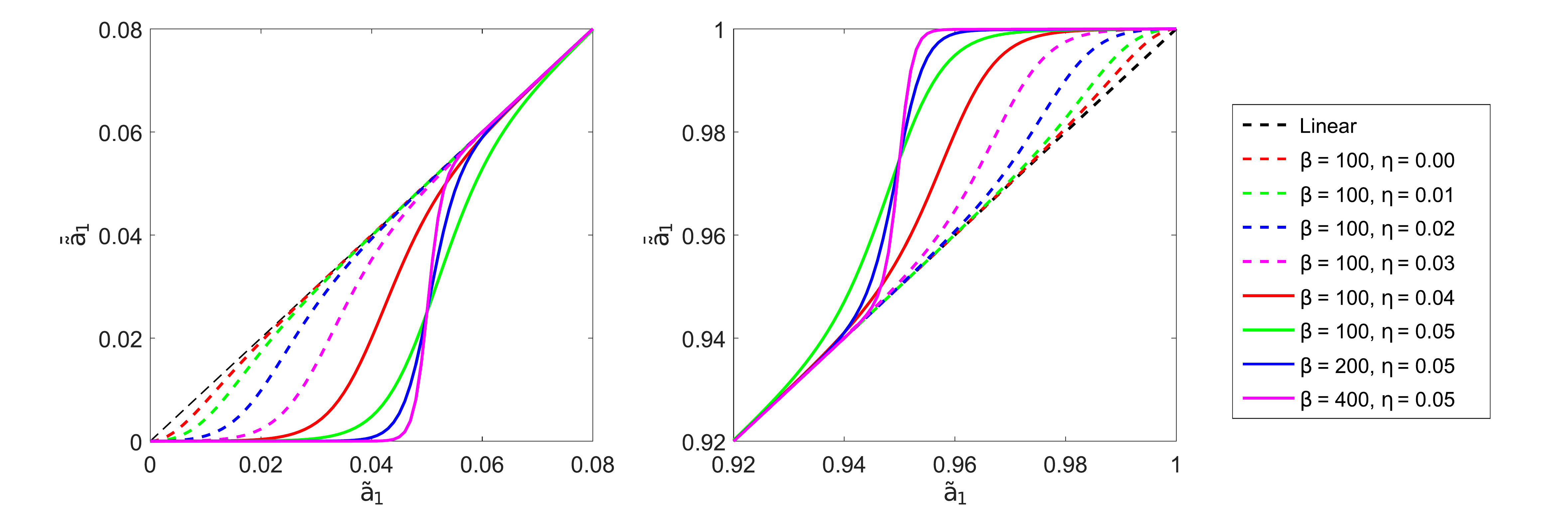}
        \caption{Interpolation scheme plotted for the intervals where the behavior is non-linear, for different values of $\eta$ and $\beta$, that follow the order of the continuation approach.}
		\label{Fig:RegWidth.1}
\end{figure}

Additionally, we want to avoid very thin and very thick layers. Instead we want the layers to be either void, completely solid or in the interval $[\eta,(1-\eta)]$, with $\eta=0.05$ used in this study. To do so, we use the same interpolation scheme as proposed in~\citep{Bib:GroenSigmund2017} that links the filtered relative layer widths $\tilde{\mu}_{i}$ to the physical relative layer widths $\bar{\tilde{\mu}}_{i}$,
\begin{equation}\label{Eq:RegWidth.1}
\begin{aligned}
\bar{\tilde{\mu}}_{i} = \tilde{\mu}_{i} \big(1-\bar{H}(\beta,(1-\eta),\tilde{\mu}_{i})\big)\bar{H}(\beta,\eta,\tilde{\mu}_{i}) + \Bigg(\frac{\beta-1}{\beta}+\frac{\tilde{\mu}_{i}}{\beta} \Bigg) \bar{H}(\beta,(1-\eta),\tilde{\mu}_{i}).
\end{aligned}
\end{equation}
Where $\bar{H}$ is the smoothed Heaviside function~\citep{Bib:WangRobust2010},
\begin{equation}\label{Eq:RegWidth.2}
\bar{H}(\beta,\eta, \tilde{\mu}_{i}) = \frac{\text{tanh}(\beta \eta) + \text{tanh}(\beta (\tilde{\mu}_{i}- \eta))}{\text{tanh}(\beta \eta) + \text{tanh}(\beta (1- \eta))},
\end{equation}
with parameter $\beta$ controlling the sharpness of the projection and $\eta$ the threshold parameter. The interpolation curves for different values of $\beta$ and $\eta$ is shown in Figure~\ref{Fig:RegWidth.1}. The order of lines in the legend shows the continuation approach that is taken, with $50$ iterations per step. This means that the material interpolation scheme begins close to a linear function, gradually $\eta$ is increased to enforce a length-scale on the relative layer widths. Finally, $\beta$ is increased to a high value to ensure that $\bar{\tilde{\mu}}_{i}$ is either $0$, $1$ or in the region $[0.05,0.95]$.

\subsection{Optimization and regularization of the microstructure orientation}
It is known that a microstructure with an elasticity tensor having orthotropic symmetry conditions is optimally aligned using the principal stress directions when a single load case problem is considered~\citep{Bib:Pedersen1990,Bib:Cowin1994,Bib:Norris2005}. It is therefore appealing to align the microstructure normals $\boldsymbol{n}^{1},\boldsymbol{n}^{2},\boldsymbol{n}^{3}$ with the eigenvectors $\boldsymbol{v}^{I},\boldsymbol{v}^{I},\boldsymbol{v}^{III}$ corresponding to the principal stresses. Unfortunately, solving the cubic equation for the 3D eigenvalue problem leads to an arbitrary order of the principal stresses as $\sigma_{I} \geq \sigma_{II} \geq \sigma_{III}$. This means that eigenvectors can swap 90 degrees when there is multiplicity in eigenvalues, in turn leading to instability in the optimization when the layer normals interchange. To circumvent these instabilities we update the orientation vectors based on the gradients w.r.t. ${\theta}_{1},{\theta}_{2}$ and ${\theta}_{3}$. 

For the de-homogenization approach to work it is very important that $\boldsymbol{n}^{1},\boldsymbol{n}^{2},\boldsymbol{n}^{3}$ are smooth and continuous throughout the design domain $\Omega$. This is unfortunately neither possible using the principal stress directions, nor using gradient-based optimization of the Euler angles. As possible solution one can regularize the orientation field using the approach by~\citet{Bib:Perle3D}; however, in our experience this may lead to misaligned members w.r.t. the load path. 
To have a de-homogenized design that performs well, we propose a computational approach in which the smooth and continuous vector fields $\tilde{\boldsymbol{n}}^{1},\tilde{\boldsymbol{n}}^{2},\tilde{\boldsymbol{n}}^{3}$ are reconstructed from $\boldsymbol{n}^{1},\boldsymbol{n}^{2},\boldsymbol{n}^{3}$. This approach, that will be described in detail in the next Section, requires that $\boldsymbol{n}^{1},\boldsymbol{n}^{2},\boldsymbol{n}^{3}$ are smoothly varying through $\Omega$ with a rotational symmetry of $\pi/2$. Hence interchanging vectors are allowed; however, large changes in frame orientation should be avoided.

\begin{figure}[h!]
\centering
\hfill
\subfloat[Orientation vectors $\boldsymbol{n}^{i}$ corresponding to face $f$.]{\includegraphics[width=0.42\textwidth]{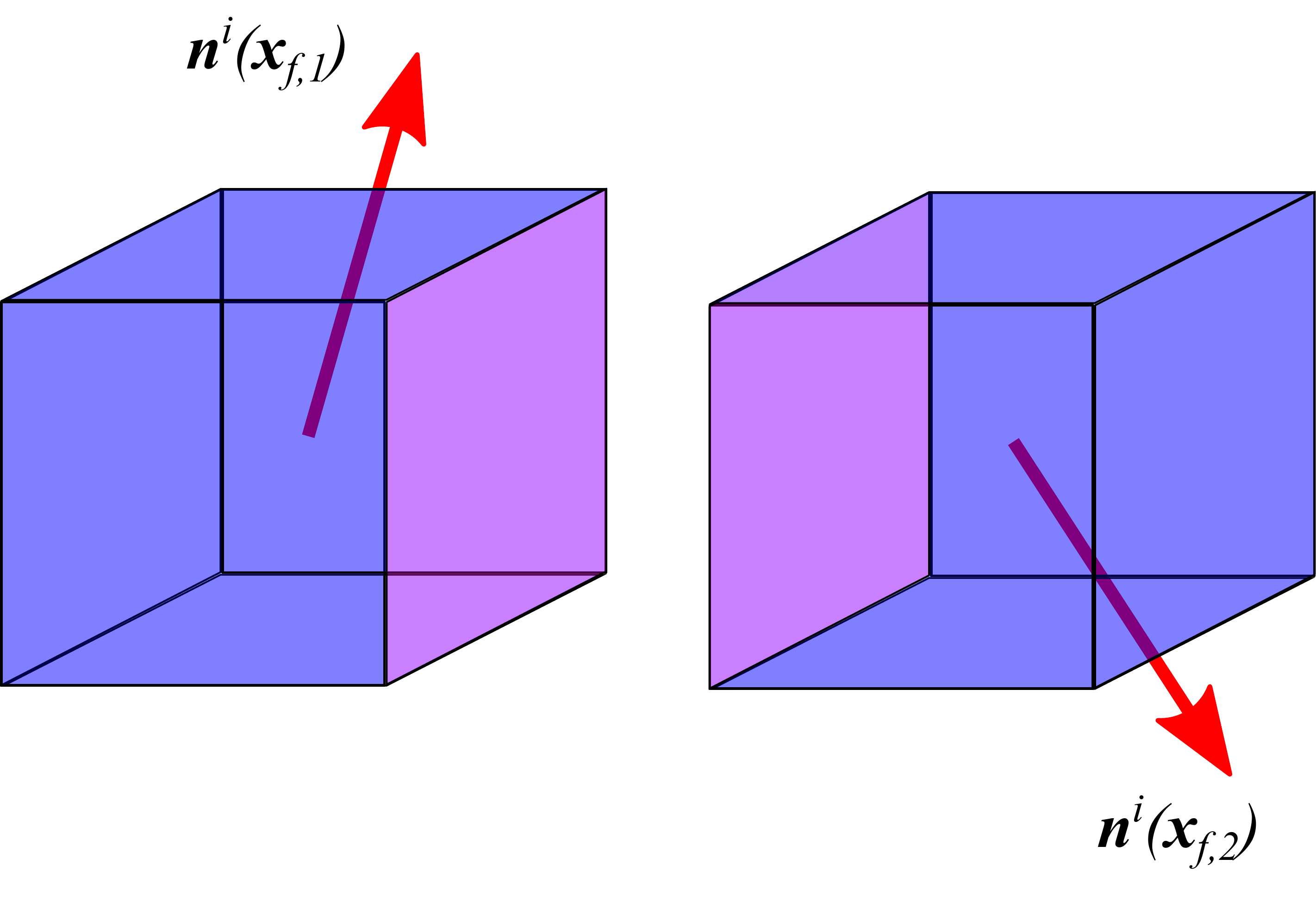}} \hfill
\subfloat[$\mathcal{P}_{fi}$ as a function of the inner product.]{\includegraphics[width=0.42\textwidth]{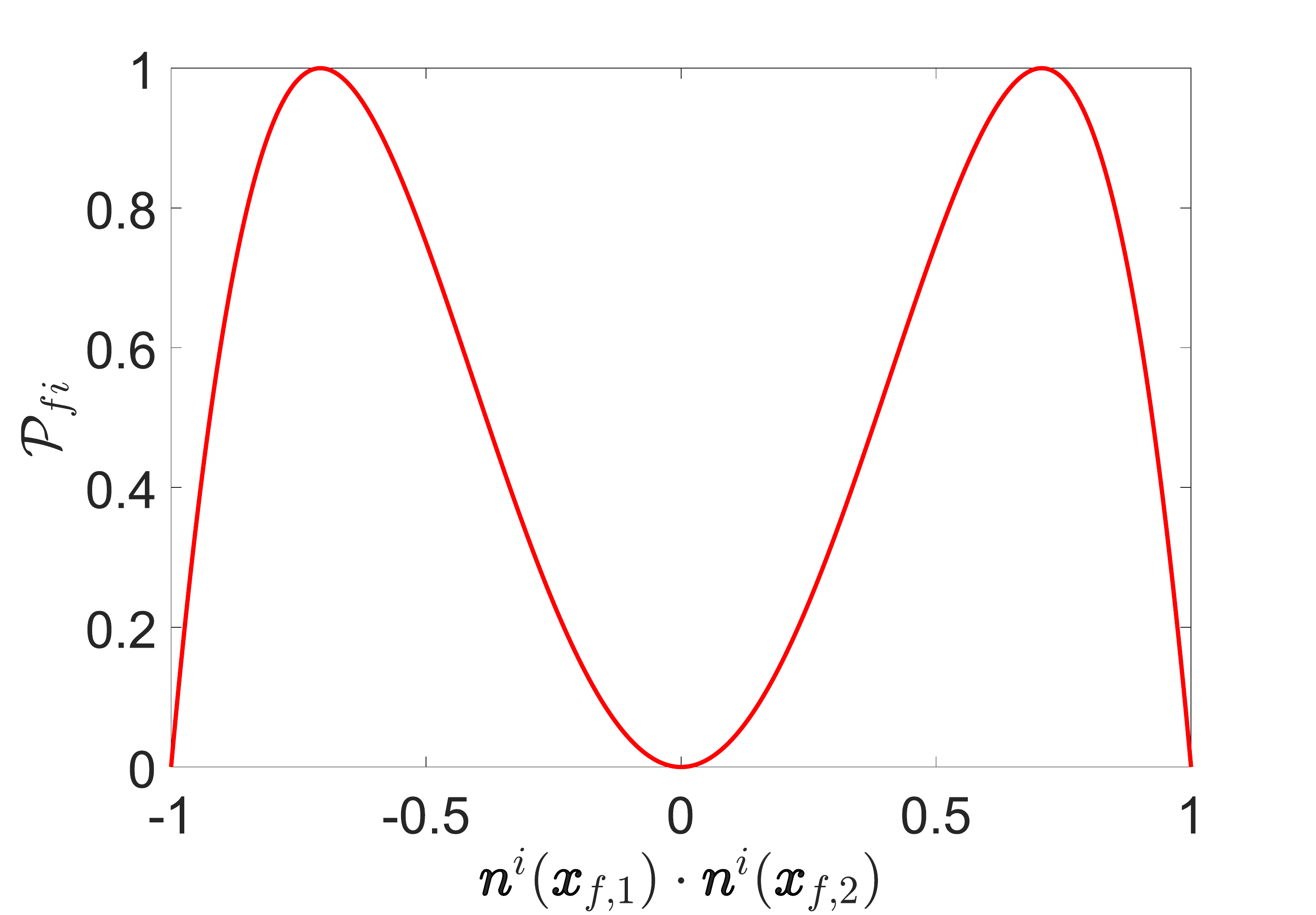}} \hfill
\caption{Visualization of the penalization of difference in orientation vectors in two elements connected by face $f$.}
\label{Fig:AngleReg.1}
\end{figure}

We perform a regularization step on the discretized mesh and consider all faces $n_{f}$ connecting two elements. A penalization function $\mathcal{P}_{fi}\in[0,1]$ is introduced that penalizes the difference between two vectors $\boldsymbol{n}^{i}(\boldsymbol{x}_{f,1})$ and $\boldsymbol{n}^{i}(\boldsymbol{x}_{f,2})$ that are connected by face $f$,
\begin{equation}\label{Eq:AngleReg.1}
\mathcal{P}_{fi} = 4 \big(\boldsymbol{n}^{i}(\boldsymbol{x}_{f,1})\cdot \boldsymbol{n}^{i}(\boldsymbol{x}_{f,2})\big)^{2} - 4 \big(\boldsymbol{n}^{i}(\boldsymbol{x}_{f,1})\cdot \boldsymbol{n}^{i}(\boldsymbol{x}_{f,2})\big)^{4}.
\end{equation}
This function has a minimum value if the vectors have the same orientation or an angle difference of $k\pi/2$, for an integer $k$, and a maximum value for a relative angle difference of $\pi/4 + k\pi/2$. To demonstrate this, consider face $f$ and corresponding normal vectors $\boldsymbol{n}^{i}(\boldsymbol{x}_{f,1})$ and $\boldsymbol{n}^{i}(\boldsymbol{x}_{f,2})$  shown in Figure~\ref{Fig:AngleReg.1}(a). The value of penalization function $\mathcal{P}_{fi}$ plotted against the inner product of the two corresponding vectors is shown in Figure~\ref{Fig:AngleReg.1}(b).

By looping over the three normal vector directions, and all faces $n_{f}$ connecting two elements, we can obtain a single regularization objective $\mathcal{F}_{\theta}$,
\begin{equation}\label{Eq:AngleReg.2}
\mathcal{F}_{\theta} = \Big( \sum_{f = 1}^{n_{f}} \sum_{i = 1}^{3} \mathcal{P}_{fi}^{q} \Big)^{1/q}.
\end{equation}
Numerical experiments have shown that a norm aggregation of $q=1$ yields the best results. Finally, it should be noted that regularization objective $\mathcal{F}_{\theta}$ can be used to augment the optimization objective or can be imposed as a constraint. The sensitivities w.r.t. Euler angles $\boldsymbol{\theta}_{1},\boldsymbol{\theta}_{2}$ and $\boldsymbol{\theta}_{3}$ can be derived analytically to allow for gradient-based optimization.

\subsection{Topology optimization problem}
The goal of the homogenization-based topology optimization problem is to minimize objective functional $\mathcal{F}$, which is a combination of the compliance $\mathcal{J}$ ($i.e.$ the external work), and regularization objective $\mathcal{F}_{\theta}$. The domain is discretized using $n_{x} \times n_{y}\times n_{z}$ tri-linear finite elements, and hence the design variables can be discretized in design vectors $\boldsymbol{\mu}_{1}, \boldsymbol{\mu}_{2}, \boldsymbol{\mu}_{3}, \boldsymbol{\theta}_{1}, \boldsymbol{\theta}_{2}$ and $\boldsymbol{\theta}_{3}$. The optimization problem is solved in a nested form, which means that for each design iteration we solve the state equation after which the design vectors are updated. Hence, the discretized optimization problem can be written as,
\begin{equation}\label{Eq:TopOpt.1}
\begin{aligned}
 & & \displaystyle \min_{\boldsymbol{\mu}_{1},\boldsymbol{\mu}_{2},\boldsymbol{\mu}_{3},\boldsymbol{\theta}_{1},\boldsymbol{\theta}_{2},\boldsymbol{\theta}_{3}} & :   \mathcal{F}(\boldsymbol{\mu}_{1},\boldsymbol{\mu}_{2},\boldsymbol{\mu}_{3},\boldsymbol{\theta}_{1},\boldsymbol{\theta}_{2},\boldsymbol{\theta}_{3},\textbf{U})= \gamma_{c} \mathcal{J}(\boldsymbol{\mu}_{1},\boldsymbol{\mu}_{2},\boldsymbol{\mu}_{3},\boldsymbol{\theta}_{1},\boldsymbol{\theta}_{2},\boldsymbol{\theta}_{3},\textbf{U}) + \gamma_{\theta} \mathcal{F}_{\theta} (\boldsymbol{\theta}_{1},\boldsymbol{\theta}_{2},\boldsymbol{\theta}_{3}),     \\
 & & \textrm{s.t.}             & :  \textbf{K}(\boldsymbol{\mu}_{1},\boldsymbol{\mu}_{2},\boldsymbol{\mu}_{3},\boldsymbol{\theta}_{1},\boldsymbol{\theta}_{2},\boldsymbol{\theta}_{3}) \textbf{U} = \textbf{F}, 								   \\ 
 & & 						     & :   \textbf{v}^{T} \boldsymbol{\rho}(\boldsymbol{\mu}_{1},\boldsymbol{\mu}_{2},\boldsymbol{\mu}_{3})  - V^{max}_{f} V\leq 0,  	\\
 & &                           & :     \textbf{0}  \leq  \boldsymbol{\mu}_{1},\boldsymbol{\mu}_{2},\boldsymbol{\mu}_{3}   \leq  \textbf{1},	\\			
 & &                           & :     \textbf{-4}\boldsymbol{\pi } \leq  \boldsymbol{\theta}_{1},\boldsymbol{\theta}_{2},\boldsymbol{\theta}_{3}   \leq  \textbf{4}\boldsymbol{\pi },				\\
\end{aligned}
\end{equation}
here $\textbf{v}$ is the vector containing the element volumes and $V^{max}_{f}$ is the maximum allowed fraction of the material in $\Omega$, with $V$ the volume of $\Omega$. $\textbf{K}$ is the stiffness matrix and vector $\textbf{F}$ describes the loads acting on the domain. We solve for the displacement vector $\textbf{U}$ using a conjugate gradient method in combination with a geometrical multigrid pre-conditioner~\citep{Bib:Amir2013}. For the design update the MATLAB implementation of the Method of Moving Asymptotes (MMA) introduced by~\citet{Bib:MMA} is used.  

As a starting guess for the layer widths we use $\boldsymbol{\mu}_{1}=\boldsymbol{\mu}_{2}=\boldsymbol{\mu}_{3}$, such that the volume constraint is exactly satisfied. The starting guess for the orientation is based on a pre-analysis using isotropic microstructures, the corresponding principal stress directions are used to determine $\boldsymbol{\theta}_{1},\boldsymbol{\theta}_{2}$ and $\boldsymbol{\theta}_{3}$. 

Finally it should be mentioned that the scaling parameters $\gamma_{c}$ and $\gamma_{\theta}$ have a large influence on the optimization procedure. $\gamma_{c}$ is based on compliance of the first analysis step $\mathcal{J}^{(1)}$, while $\gamma_{\theta}$ is based on the regularization objective for the starting guess $\mathcal{F}^{(1)}_{\theta}$, such that,
\begin{equation} \label{Eq:TopOpt.1}
\begin{aligned}
&& \gamma_{c} = \frac{1}{\mathcal{J}^{(1)}}, && \gamma_{\theta} = \frac{1}{2 \mathcal{F}^{(1)} _{\theta}}. &&
\end{aligned}
\end{equation}

\subsection{Numerical examples}
\begin{figure}[ht!]
\centering
\includegraphics[width=0.9\textwidth]{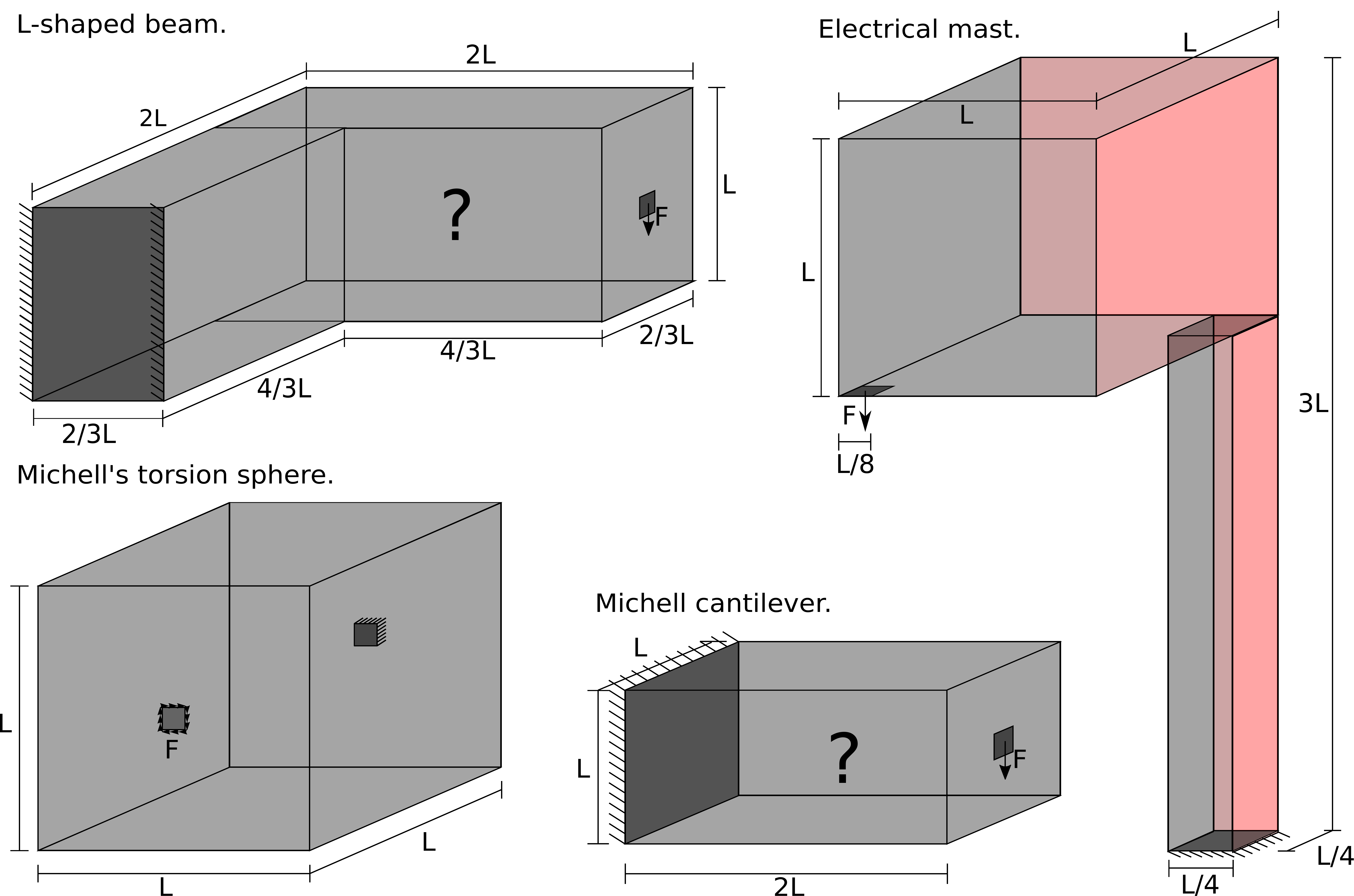}
        \caption{Dimensions and boundary conditions for the four numerical examples used in this work.}
		\label{Fig:TOEx.1}
\end{figure}

In this paper we will use four different examples, comprising the Michell cantilever, Michell's torsion sphere, an electrical mast example and the L-shaped beam all shown in Figure~\ref{Fig:TOEx.1}.

For the Michell cantilever the load is applied in a distributed sense over a patch of $L/6\times L/6$ with a magnitude of $36/L^{2}$. Furthermore, this patch of elements is set to solid with a depth of $L/24$ into the domain. For the torsion sphere both the load and the Dirichlet boundary condition are applied on a square with dimensions $L/12\times L/12$. The load is applied as a line load along the boundary of this square using a magnitude of 3/L. Finally, there are solid elements at both boundary conditions. The electrical mast example is inspired by~\citet{Bib:Perle3D}. Like them, we only model a fourth of the full structure, hence the red shaded boundaries represent symmetry conditions. The load is applied in a distributed sense over a patch of $L/8\times L/8$ with a magnitude of $64/L^{2}$. The L-shaped design domain, consists of passive elements to show a torsion bending coupling. The load is applied in a distributed sense over a square patch of solid materials of $L/6\times L/6$ with a magnitude of $36/L^{2}$. Finally, it has to be mentioned that all examples are obtained using a Young's modulus for the stiff material $E^{+} = 1$, and $E^{-} = 10^-{3}$ to represent the weak material in the rank-3 microstructures, since lower values result in a much slower convergence. Furthermore, we use a length $L = 1$ and a maximum allowed volume fraction $V^{max}_{f} = 0.1$.

To demonstrate the effect between the angle optimization and regularization methods we consider the electrical mast example. The structure is optimized on two different mesh sizes, and corresponding compliance values on the coarse optimization mesh $\mathcal{J}^{c}$ are shown in Table~\ref{Tab:TOEx.1}. We demonstrate the difference between gradient-based alignment of the microstructure and the use of a stress-based alignment procedure. Furthermore, we show the effect of the starting guess for the microstructure orientation, which can be either based on stress directions in a pre-analysis or by setting the Euler angles to zero. Finally, we show the effect of regularization on the angles and the regularization scheme that avoids thin features in the relative widths of the layers. The method used in this work is denoted in boldface.
\begin{table}[ht!]
\centering
\caption{Compliance values $\mathcal{J}^{c}$ for the electrical mast example, optimized using different discretizations, alignment methods, starting guesses and regularization schemes. The method used in this work is denoted in boldface.}
\label{Tab:TOEx.1}
\centering
\setlength{\arrayrulewidth}{1pt} 
\begin{tabular}{ccccc}
 \hline
\textbf{Mesh size} & \textbf{alignment method} & \textbf{Starting orientation}& \textbf{Regularization method} & $\mathcal{J}^{c}$ \\ \hline
$\mathbf{24\times24\times72}$& \textbf{gradient-based} &	\textbf{principal directions} &	\textbf{Both widths and angles} & $\mathbf{98.23}$ \\
$24\times24\times72$& gradient-based &	principal directions &	Only on widths & 97.12 \\
$24\times24\times72$& gradient-based &	principal directions &	Only on angles & 99.17 \\
$24\times24\times72$& gradient-based &	principal directions &	No regularization & 97.17 \\
$24\times24\times72$& gradient-based &	$\boldsymbol{\theta}_{1}= \boldsymbol{\theta}_{2} = \boldsymbol{\theta}_{3} = \boldsymbol{0}$&	Both widths and angles & 100.10 \\
$24\times24\times72$& gradient-based &	$\boldsymbol{\theta}_{1}= \boldsymbol{\theta}_{2} = \boldsymbol{\theta}_{3} = \boldsymbol{0}$ &	Only on widths & 97.39 \\
$24\times24\times72$& stress-based &	principal directions &	Only on widths & 97.69 \\ 
\hline 
$\mathbf{48\times48\times144}$& \textbf{gradient-based} &	\textbf{principal directions} &	\textbf{Both widths and angles} & $\mathbf{98.03}$ \\
$48\times48\times144$& gradient-based &	principal directions &	Only on widths & 97.44 \\
$48\times48\times144$& gradient-based &	principal directions &	Only on angles & 99.63 \\
$48\times48\times144$& gradient-based &	principal directions &	No regularization & 98.43 \\
$48\times48\times144$& gradient-based &	$\boldsymbol{\theta}_{1}= \boldsymbol{\theta}_{2} = \boldsymbol{\theta}_{3} = \boldsymbol{0}$&	Both widths and angles & 102.13 \\
$48\times48\times144$& gradient-based &	$\boldsymbol{\theta}_{1}= \boldsymbol{\theta}_{2} = \boldsymbol{\theta}_{3} = \boldsymbol{0}$ &	Only on angles & 98.43 \\
$48\times48\times144$& stress-based &	principal directions &	Only on widths & 97.57 \\ 
\hline
 \end{tabular}
\end{table}

As can be seen, the best results are obtained using gradient-based optimization of the orientation angles in combination with the principal directions as starting guess. In general the effect of adding the projection scheme (see Figure~\ref{Fig:RegWidth.1}) to avoid thin relative widths does not increase the compliance, in fact it improves the performance. Furthermore, the regularization of the angles only reduces the performance by $1-2\%$; however, it ensures structures that can be de-homogenized as will be discussed in the next Section.

\section{De-homogenization method}
\label{Sec:DeHomo}
A design with rank-3 microstructures can be approximated on single scale using an approach similar to the one presented in~\citep{Bib:GroenSigmund2017,Bib:GroenWuSigmund2019}.
However, some extra steps need to be taken, since the vector fields that describe the layer normals should be smooth and continuous. First of all, it is not directly possible to interchange the layer normals $\boldsymbol{n}^{i}$, $\boldsymbol{n}^{j}$ and the physical relative layer widths $\bar{\tilde{\mu}}_{i}$ and $\bar{\tilde{\mu}}_{j}$ for $i\neq j$, due to the hierarchy contained in $\bar{\tilde{\mu}}_{i}$ Hence, before the three layers are sorted to be smooth and continuous throughout domain $\Omega$ it is important that this hierarchy is taken into account. Afterwards, we have three orthogonal vector fields that are aligned up to $\pi/2$. However, we should note that the normal vectors $\boldsymbol{n}^{1},\boldsymbol{n}^{2},\boldsymbol{n}^{3}$ describing the microstructure orientation are rotationally symmetric up to $\pi$, $i.e.$ using $-\boldsymbol{n}^{1}$ instead of $\boldsymbol{n}^{1}$ results in the same microstructures. Hence, we have a 3-dimensional 6-direction field from which 3 orthogonal, smooth and continuous 1-direction fields $\tilde{\boldsymbol{n}}^{1},\tilde{\boldsymbol{n}}^{2},\tilde{\boldsymbol{n}}^{3}$ have to be extracted.

\subsection{Single scale interpretation of a rank-3 microstructure}
As discussed in~\citet{Bib:TraffSigmundGroen2019} for the 2D case, it is possible to approximate a multi-scale rank-3 laminate with small loss in performance on a single scale. Here, we extend the idea to orthogonal rank-3 microstructures in 3D. To do so we make use of the relative layer contributions $p_{i}$. These layer contributions are linked to the physical relative layer width $\bar{\tilde{\mu}}_{i}$ using,
\begin{equation}\label{Eq:Sort.1}
\begin{aligned}
p_{1} &= \frac{\bar{\tilde{\mu}}_{1}}{\rho}, \\
p_{2} &= \frac{(1-\bar{\tilde{\mu}}_{1})\bar{\tilde{\mu}}_{2}}{\rho}, \\
p_{3} &= \frac{(1-\bar{\tilde{\mu}}_{1})(1-\bar{\tilde{\mu}}_{2})\bar{\tilde{\mu}}_{3}}{\rho}. \\
\end{aligned}
\end{equation}
In a subsequent step we can obtain the single scale layer widths $w_{i} = \alpha p_{i}$, with $\alpha >0 $ a scaling factor such that the following equation holds,
\begin{equation}\label{Eq:Sort.2}
\rho = \alpha (p_{1} +p_{2} + p_{3}) - \alpha^{2} (p_{1}p_{2} + p_{1}p_{3} + p_{2}p_{3}) + \alpha^{3} p_{1}p_{2}p_{3}.
\end{equation}

\subsection{Obtaining smooth and continuous vector fields}
Consider the 6-direction vector field consisting of layer normals $ \pm\boldsymbol{n}^{1},\pm\boldsymbol{n}^{2},\pm\boldsymbol{n}^{3}$ and layer widths $w_{1}, w_{2}, w_{3}$. From this we want to extract three smooth and continuous 1-direction fields $\tilde{\boldsymbol{n}}^{1},\tilde{\boldsymbol{n}}^{2},\tilde{\boldsymbol{n}}^{3}$ with sorted widths $\tilde{w}_{1}$,$\tilde{w}_{2}$, and $\tilde{w}_{3}$. Where the normal vectors $\tilde{\boldsymbol{n}}^{i}$ are smooth and continuous throughout the domain.

Similar to the method described by~\citet{Bib:Perle3D} we require that the vector fields used for the de-homogenization are free from singularities. The effect of singularities is that if you trace a curve around them and follow a given vector as you move along this curve, then the vector will be rotated when you come full circle. A more detailed explanation about vector fields, direction fields, singularities and possible treatments can be found in \citep{Bib:Quadcover2007,Bib:Nieser2011,Bib:Vaxman2016}. 

Fortunately, we observe that singularities tend to occur in (nearly) void regions outside of the mechanical structure, and this points to an effective way of computing the 1-direction fields such that the singularities do not affect our results. We observe that in the absence of singularities, we can simply propagate a consistent choice of vectors for our 1-direction fields from an initial element. If we propagate in such a way that we visit elements which might have singularities ($i.e.$ nearly void elements) only \textit{after} we have visited all of the elements pertinent to the mechanical structure, then we could stop once all significantly non-void elements have been visited and it will not be possible to draw a loop containing a singularity in this region. Clearly, this does not protect us from singularities inside the mechanical structure (regardless of density) but these seem to occur only for specific boundary conditions.

Informed by the observations above, we have designed a front propagation approach that visits all elements in density sorted order. Initially, we take a starting element with all layers widths $w_{i}\in[0.05,0.95]$ and we set $\tilde{\boldsymbol{n}}^{1}=\boldsymbol{n}^{1}, \tilde{\boldsymbol{n}}^{2}=\boldsymbol{n}^{2}, \tilde{\boldsymbol{n}}^{3}=\boldsymbol{n}^{3}$, hence the widths follow such that $\tilde{w}_{1} = w_{1}$, $\tilde{w}_{2} = w_{2}$, and $\tilde{w}_{3} = w_{3}$. Then we add its neighbors to a priority queue, $\boldsymbol{Q}$. The priority is given by $\lvert 0.5-\boldsymbol{\rho} \rvert$ where smaller values correspond to higher priority. Subsequently, when we take an element out of the queue, we fix its 1-direction fields in a way discussed below, and add its non-visited neighbors to $\boldsymbol{Q}$. Finally, we mark this element as visited in a vector $\boldsymbol{V}$. This procedure leads to a traversal of all elements in order of density closest to $0.5$, but in a spatially contiguous fashion. If there is a singularity in the void domain outside the mechanical structure, it will not influence the direction fields computed by this approach. 

If we take an element $e$ out of $\boldsymbol{Q}$, we find the right handed frame $\tilde{\boldsymbol{F}}_{e}$ that describes the layer widths,
\begin{equation}\label{Eq:Sort.3}
\tilde{\boldsymbol{F}}_{e} = \begin{bmatrix} \tilde{\boldsymbol{n}}^{1}(\boldsymbol{x}_{e}) & \tilde{\boldsymbol{n}}^{2}(\boldsymbol{x}_{e}) & \tilde{\boldsymbol{n}}^{3}(\boldsymbol{x}_{e})\end{bmatrix}.
\end{equation}
There are $j=24$ possible frame orientations $\boldsymbol{F}^{j}_{e}$ that have to be tested to find the best $\tilde{\boldsymbol{F}}_{e}$. To do this, we identify the number of neighbor elements $n^{n}$ that already have been visited, and identify the possible rotation matrix $\boldsymbol{R}_{e,i}^{j}$ with respect to each of the already defined frame orientations of neighbor elements $i\in n^{n}$,
\begin{equation}\label{Eq:Sort.4}
\boldsymbol{R}_{e,i}^{j} = \tilde{\boldsymbol{F}}_{i}^{T} \boldsymbol{F}^{j}_{e}.
\end{equation}
The corresponding orientation angle $\psi_{e,i,j}$ that defines the frame orientation can be calculated as,
\begin{equation}\label{Eq:Sort.5}
\lvert \psi_{e,i,j} \rvert = \text{arccos}\Bigg( \frac{\text{trace}(\boldsymbol{R}_{e,i}^{j}) - 1}{2} \Bigg),
\end{equation}
Hence, $\psi_{e,i,j}=0$ would mean that the frame in $e$ coincides with the frame in $i$ for a given possibility $j$. The best orientation follows as,
\begin{equation}\label{Eq:Sort.6}
\begin{aligned}
\tilde{\boldsymbol{F}}_{e} =\boldsymbol{F}_{e}^{k}, && \text{for} && k  = \text{arg} \min_{j = 1,\ldots, 24} \sum_{i} \lvert \psi_{e,i,j} \rvert.
\end{aligned}
\end{equation}
Once, we have the sorted vectors for element $e$, $\tilde{\boldsymbol{n}}^{1},\tilde{\boldsymbol{n}}^{2},\tilde{\boldsymbol{n}}^{3}$ that describe frame $\tilde{\boldsymbol{F}}_{e}$ we store the corresponding widths $\tilde{w}_{i}$. Subsequently we remove element $e$ from the queue and mark it as visited; furthermore, we add the unvisited neighbors of element $e$ to the queue and sort again based on density. Subsequently, we take the element with the highest priority out of the queue and perform the same procedure again. This process is repeated until we have three smooth 1-direction fields.

\subsection{De-homogenization of multi-scale designs}
Similar to the work presented in~\citep{Bib:GroenSigmund2017} we need to calculate a mapping field $\phi_{i}$ for each of the three layers. Using this mapping function we can create an implicit geometry description $\tilde{\rho}_{i}(\boldsymbol{x})$ of the $i$-$th$ layer,
\begin{equation}\label{Eq:Proj.1}
\tilde{\rho}_{i}(\boldsymbol{x}) = H\Bigg( \big(\frac{1}{2} + \frac{1}{2}\mathcal{S}\left\{ P_{i} \phi_{i}(\boldsymbol{x}) \right\} \big)  - \tilde{w}_{i} (\boldsymbol{x})\Bigg).
\end{equation}
Here $H$ is the Heaviside function and $\mathcal{S}\in[-1,1]$ corresponds to the \texttt{sawtooth} function in MATLAB. Furthermore, $P_{i}$ is a periodicity scaling, which as will be discussed later, depends on mapping function $\phi_{i}$. The three implicit geometry functions for each layer can be combined to create an implicit geometry description of the de-homogenized structure,
\begin{equation}\label{Eq:Proj.2}
\tilde{\rho}(\boldsymbol{x}) = \min \left\{ 1,\sum_{i=1}^{3} \tilde{\rho}_{i}(\boldsymbol{x})\right\}.
\end{equation}
Since small widths are avoided using the continuation scheme presented in Figure~\ref{Fig:RegWidth.1} we only need an accurate description of $\phi_{i}$ in $\tilde{\Omega}_{i}$,
\begin{equation}\label{Eq:Proj.3}
\begin{aligned}
\boldsymbol{x} \in \tilde{\Omega}_{i}  &\quad \text{if }\quad  \tilde{w}_{i}(\boldsymbol{x}) > 0.01 \text{ and }  \rho(\boldsymbol{x}) < 0.99.
\end{aligned}
\end{equation}
To solve for $\phi_{i}$ we solve the following least-squares problem,
\begin{equation}\label{Eq:Proj.4}
\begin{aligned}
\displaystyle \min_{\phi_{i}(\boldsymbol{x})} & :  \mathcal{I}(\phi_{i}(\boldsymbol{x}))		      =  \frac{1}{2}\int_{\Omega} \alpha^{i}_{1}(\boldsymbol{x})\left \Vert \nabla \phi_{i}(\boldsymbol{x})  -  \tilde{\boldsymbol{n}}^{i}(\boldsymbol{x}) \right \Vert ^{2} \textnormal{d}\Omega,&  &  \\
\textrm{s.t.}             		 & : \alpha^{i}_{2}(\boldsymbol{x}) \nabla \phi_{i} (\boldsymbol{x})\cdot \tilde{\boldsymbol{t}}^{i,1}(\boldsymbol{x}) = 0, &  & \\ 
\textrm{s.t.}             		 & : \alpha^{i}_{2}(\boldsymbol{x}) \nabla \phi_{i} (\boldsymbol{x})\cdot \tilde{\boldsymbol{t}}^{i,2}(\boldsymbol{x}) = 0. &  & \\ 
 \end{aligned}
\end{equation}
The domain is split into three parts, which dictate the weights on the objective $\alpha^{i}_{1}$ and the weights on the constraints $\alpha^{i}_{2}$,
\begin{equation}\label{Eq:Proj.5}
\begin{aligned}
 \alpha^{i}_1(\boldsymbol{x}) =
  \begin{cases}
    0.01       & \quad \text{if }\quad \tilde{w}_{i}(\boldsymbol{x}) < 0.01, \\
    0.1  & \quad \text{if } \quad \rho(\boldsymbol{x})>0.99, \\
	1 & \quad \text{if } \quad \boldsymbol{x}\in \tilde{\Omega}_{i}.\\ 
  \end{cases}
  &,&
   \alpha^{i}_2(\boldsymbol{x}) =
  \begin{cases}
    0       & \quad \text{if }\quad  \tilde{w}_{i}(\boldsymbol{x})< 0.01, \\
    0  & \quad \text{if } \quad \rho(\boldsymbol{x}) >0.99, \\
	1 & \quad \text{if } \quad \boldsymbol{x}\in \tilde{\Omega}_{i}.\\ 
  \end{cases}
  \end{aligned}
\end{equation}
The term $\alpha^{i}_{1}$ is introduced to relax the requirements for $\phi_{i}$ in regions that are either solid or void, where the low values still ensure some regularization to the lattice spacing. Furthermore, the term $\alpha^{i}_{2}$ is used to turn off exact angular enforcement in these regions. Numerically, we solve the above mentioned problem using a finite element approach, where the constraints are enforced in an augmented setting using a penalty parameter $\gamma_{\phi}$. Furthermore, the complete sequence of topology optimization, creating mapping fields, and creating an implicit geometry description can be solved in a multi-scale manner. It is known that homogenization-based topology optimization can be performed on a relatively coarse mesh $\mathcal{T}^{c}$, while it can still contain a lot of details. The implicit geometry function $\tilde{\rho}$ is a continuous description of the de-homogenized shape as long as the mapping functions $\phi_{i}$ and widths $\tilde{w}_{i}$ have a continuous description. For practical purposes $\tilde{\rho}$ is evaluated using a discrete number of points; however, on a fine mesh $\mathcal{T}^{f}$, such that $h^{f}  \leq h^{c}/20$.

Finally, it should be mentioned that we can impose an average layer spacing $\varepsilon$, which can be interpreted as the unit-cell size. To do so, we define the periodicity scaling parameter $P_{i}$ based on the average lattice spacing in the domain of interest $\tilde{\Omega}_{i}$,
\begin{equation}\label{Eq:Proj.6}
{P}_{i} = \frac{2\pi}{\epsilon}\frac {\int_{\tilde{\Omega}_{i}} \text{d}\tilde{\Omega}_{i}}{\int_{\tilde{\Omega}_{i}} ||\nabla \phi_{i}(\boldsymbol{x})||  \text{d}\tilde{\Omega}_{i}}.
\end{equation}

\subsection{Verification and clean up of de-homogenized designs}
The de-homogenized designs can be interpolated on a very fine voxel grid, since an interpolation basis can be made for mapping functions $\phi_{i}$ and widths $\tilde{w}_i$. Similar to the 2D case (see~\citet{Bib:GroenSigmund2017}) we can identify: 1) de-homogenized layers that consist of thin widths, 2) regions that do not carry load and 3) disconnected patches of material. 

De-homogenized layers consisting of thin widths can be identified by evaluating the local spacing $\lambda_{i}$, corresponding to layer $i$, 
\begin{equation}\label{Eq:LS.1}
\lambda_{i}(\boldsymbol{x}) = \frac{2\pi}{P_{i}\lVert \nabla \phi_{i}(\boldsymbol{x}) \rVert} .
\end{equation}
This local spacing can be used to get a description of the actual feature size on the solid $f_{i}$ for layer $i$. 
\begin{equation}\label{Eq:LS.2}
 f_{i}(\boldsymbol{x}) =  \tilde{w}_{i}(\boldsymbol{x}) \lambda_{i}(\boldsymbol{x}).
\end{equation} 
Similar to~\citet{Bib:GroenSigmund2017}, we add a minimum feature size $f_{min}$ to the solid at places where this feature size was violated. To do so we obtain a modified local width $\tilde{w}_{i}^{*}$ in cases where the feature size on the solid is violated, such that
\begin{equation}\label{Eq:LS.3}
 \tilde{w}_{i}^{*}(\boldsymbol{x})  =   \begin{cases}
    \tilde{w}_{i}(\boldsymbol{x}),      & \quad \text{if }\quad  f_{i}(\boldsymbol{x}) \geq f_{min}, \\
	\frac{f_{min}}{\lambda_{i}(\boldsymbol{x})} ,     & \quad \text{if }\quad  f_{i}(\boldsymbol{x}) < f_{min} .\\
  \end{cases}
\end{equation}
This new width $\tilde{w}_{i}^{*}$ is then used in Equation~\ref{Eq:Proj.2}. However, this means that the volume of the de-homogenized shape is slightly increased. This can be alleviated by proportionally removing material in the rest of the domain; however, this option is not considered in this study.

\begin{figure}[h!]
\centering
\subfloat[With floating material.]{\includegraphics[width=0.330\textwidth]{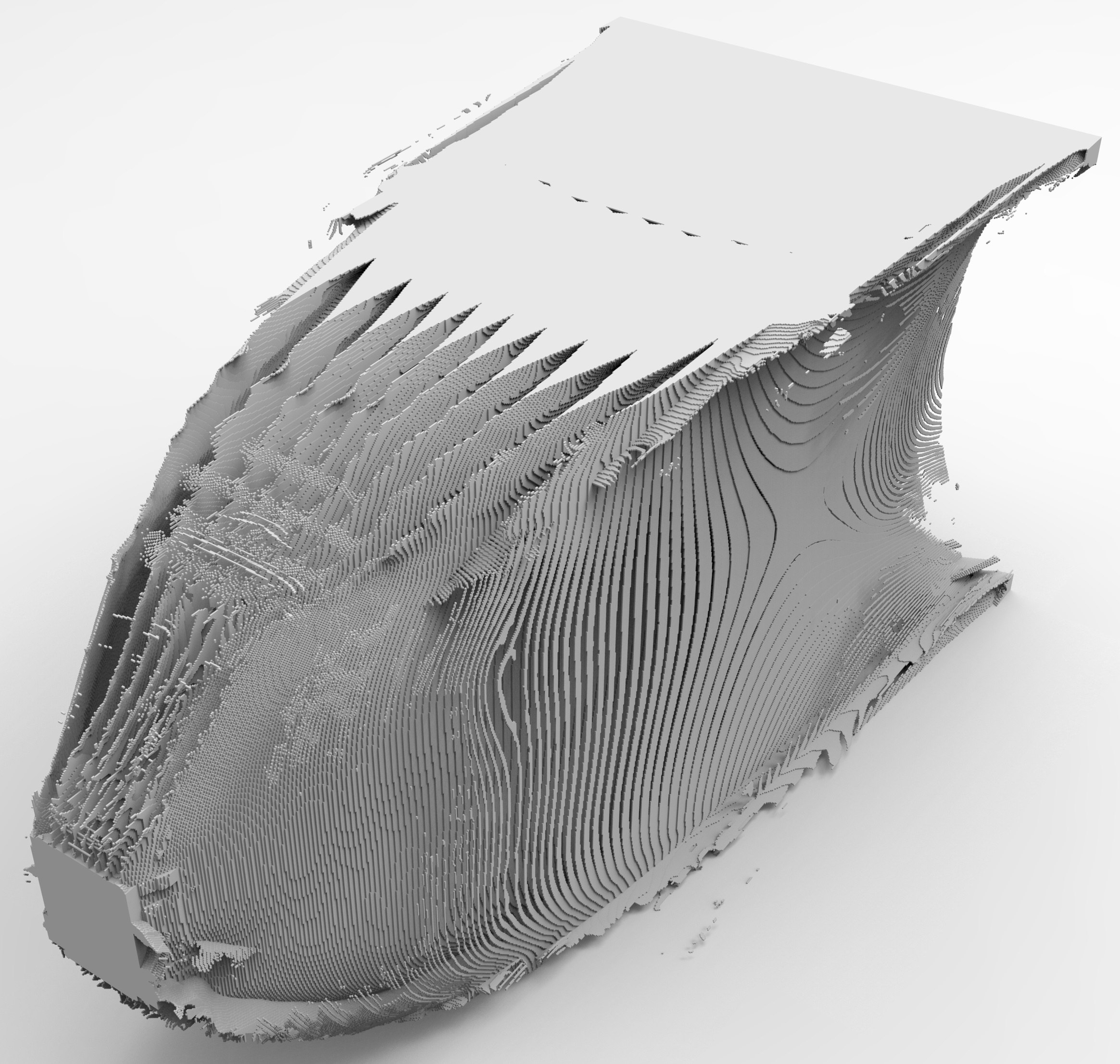}} 
\subfloat[De-homogenized design.]{\includegraphics[width=0.330\textwidth]{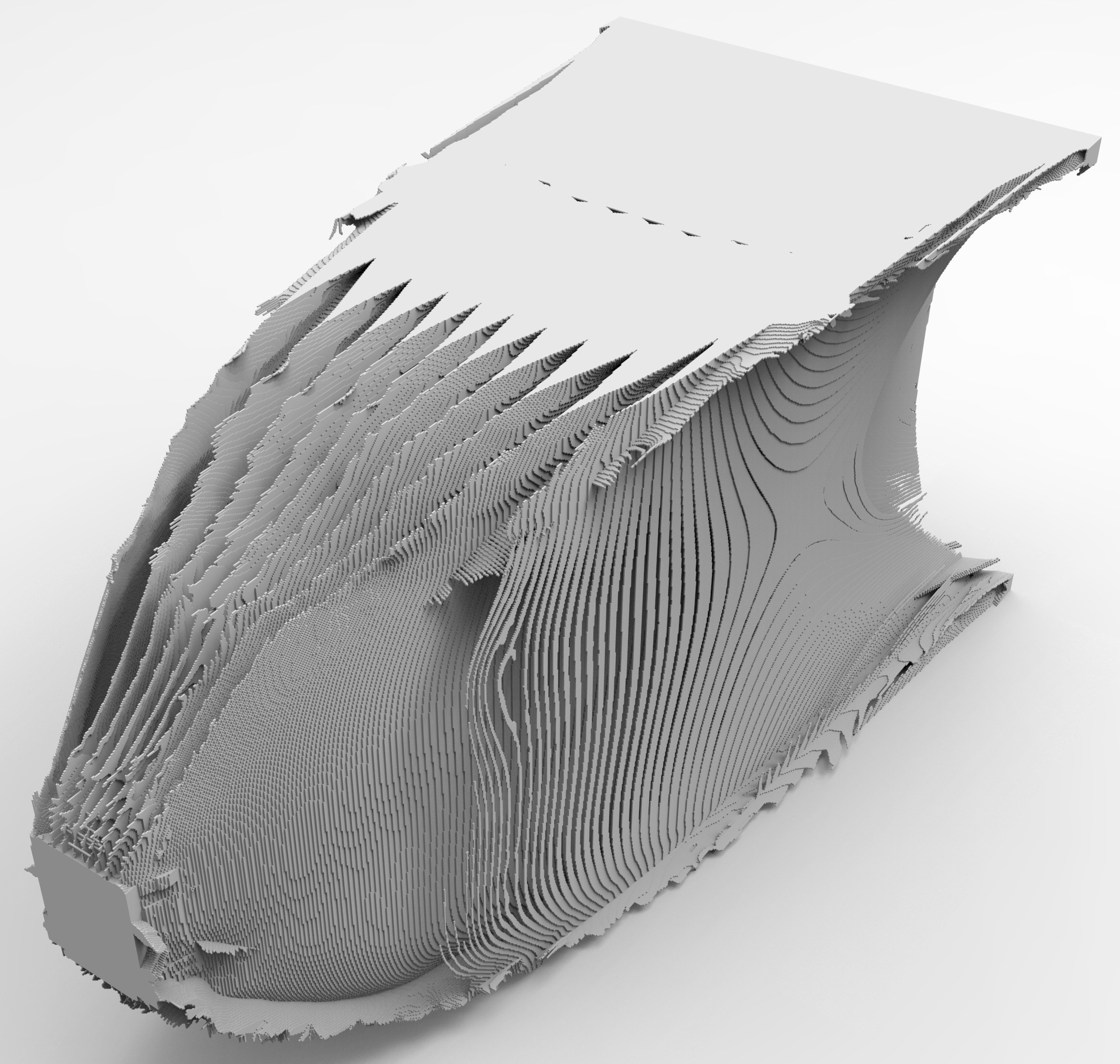}} 
\subfloat[After post-processing.]{\includegraphics[width=0.330\textwidth]{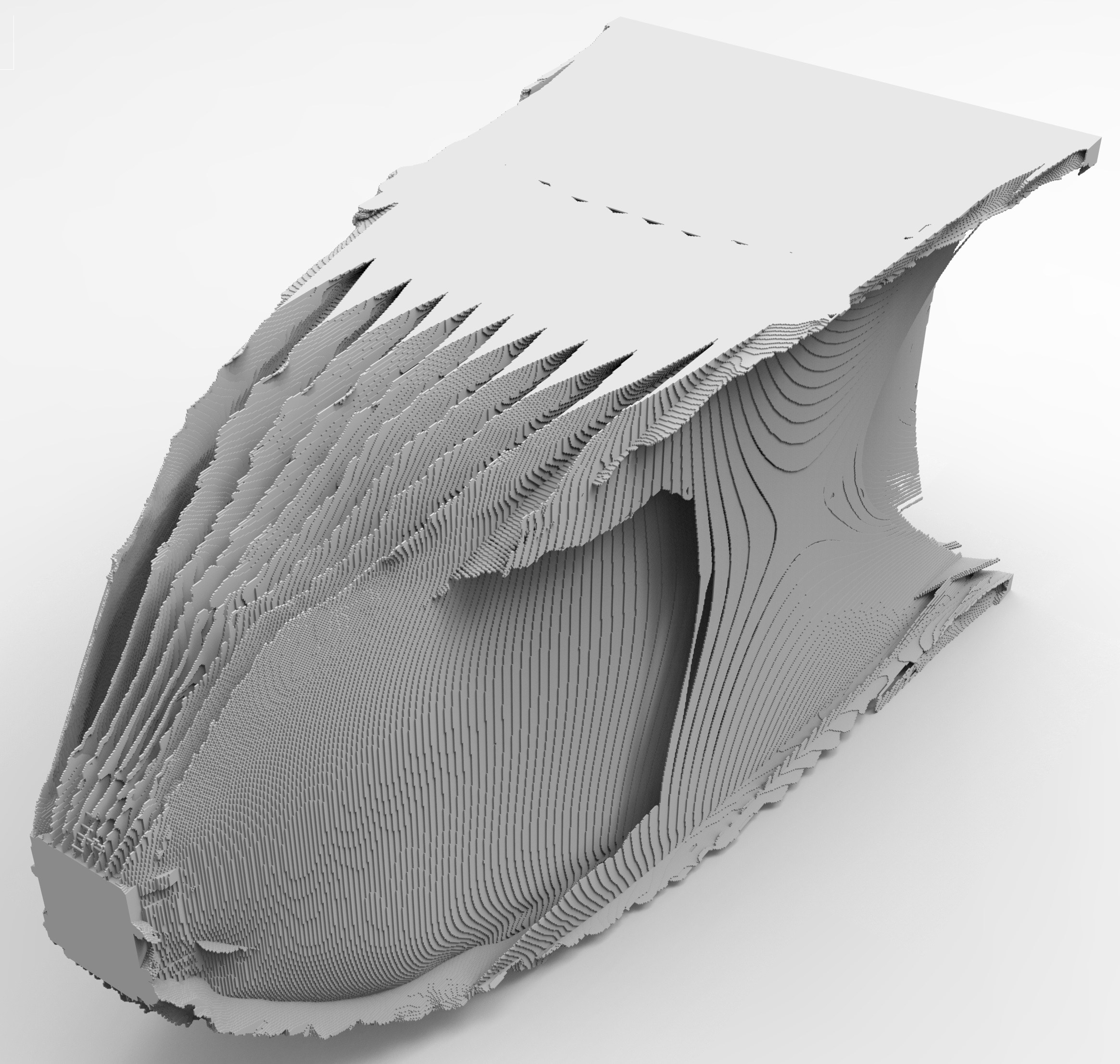}}
\caption{A Michell cantilever de-homogenized on a fine mesh of $960\times480\times480$ using $\varepsilon=30~h^{f}$ and no minimum length-scale enforcement.}
\label{Fig:Verification.1}
\end{figure}

An example of the Michell cantilever evaluated on a mesh of $960\times480\times480$ elements is shown in Figure~\ref{Fig:Verification.1}(a) where it can be seen that there are disconnected patches of material. These patches of material can be identified using the in-built connected component labeling algorithm in MATLAB ($i.e.$ \texttt{bwconncomp}), which can identify the separate components of solid material. By only keeping the component with the largest number of solid voxels, the disconnected patches are removed as can be seen in Figure~\ref{Fig:Verification.1}(b). Unfortunately, this scheme does not take care of solid regions that do not carry any load. To remove these solid elements, we make use of the simple iterative update scheme as proposed in~\citep{Bib:GroenSigmund2017}. First we perform a fine-scale finite element analysis using a slightly modified version of the publicly available topology optimization code using the PETSc framework~\citet{Bib:TopOptPETSc}. Second, the solid elements that have a strain energy density $\mathcal{E}$ lower than $10^{-2.5}$ of the mean strain energy density $\bar{\mathcal{E}}$ are set to void.  Third, to make sure that the length-scale $f_{min}$ is still satisfied after each iteration, an open-close filter operation~\citep{Bib:Sigmund2007} is applied. These steps are then repeated until no changes are made. The final design for the Michell cantilever can be seen in Figures~\ref{Fig:Verification.1}(c).

The combination of removal of solid elements with a low strain energy density, followed by the open-close filter operation, generally convergences within $5$-$10$ iterations. Besides a clean and well-connected design, the performance of the design on the fine-scale $\mathcal{J}^{f}$ is immediately known. Furthermore, it has to be noted that for the fine-scale analysis void is modeled using a Young's modulus $E^{-}=10^{-9}E^{+}$.

\section{Numerical examples}
\label{Sec:NumExp}
To demonstrate that our approach requires moderate computational resources, we perform both the homogenization-based topology optimization step and the de-homogenization step on a modern workstation PC using a single core MATLAB code. To be more specific, the PC uses Ubuntu 16.04.6 as operating system, contains an Intel Xeon Platinum 8160 processor, with $64$GB RAM memory. Although the homogenization-based designs can be evaluated on an infinitely fine voxel grid, we have been restricted by memory size to examples in the range of 200 million voxels.
These large-scale designs have been verified on the DTU Sophia cluster using 100 nodes each containing 2 AMD EPYC 7351 processors and 128GB RAM memory. Hence, a total of $3200$ cores was used.

\subsection{Homogenization-based topology optimization}
The first step in the procedure is the homogenization-based topology optimization. All examples shown in Figure~\ref{Fig:TOEx.1} are optimized on differently discretized coarse meshes $\mathcal{T}^{c}$. The corresponding compliance values on the coarse optimization meshes $\mathcal{J}^{c}$, the volume fraction $V_{f}^{c}$, number of iterations $n_{iter}$ and the total time used for the optimization $T^{c}$ are shown in Table~\ref{Tab:ExpHbased.1}.

\begin{table}[ht!]
\centering
\caption{Compliance $\mathcal{J}^{c}$, volume fraction $V_{f}^{c}$, the number of iterations $n_{iter}$, and run time $T^{c}$ for different optimization examples optimized using homogenization-based topology optimization. Reference compliance values $\mathcal{J}^{ref}$ obtained using density-based topology optimization are shown as well.}
\label{Tab:ExpHbased.1}
\centering
\setlength{\arrayrulewidth}{1pt} 
\begin{tabular}{ccccccc}
 \hline
 \textbf{Example} & $\mathcal{T}^{c}$ & $\mathcal{J}^{c}$ & $\mathcal{J}^{ref}$ &  $V_{f}^{c}$ & $n_{iter}$ & $T^{c}~[hh:mm:ss]$ \\ \hline
 Michell cantilever & $48\times24\times24$ & $227.89$ & $269.763$& $0.100$  & 400& 01:23:51 \\
 Michell cantilever & $96\times48\times48$ & $226.68$ & $265.52$& $0.100$  & 400& 09:45:35 \\
 Michell's torsion sphere & $48\times48\times48$ & $12.33$ & $12.51$ & $0.100$  & 400& 04:28:35 \\
 Michell's torsion sphere & $72\times72\times72$ & $14.00$ & $14.14$ & $0.100$  & $400$  & 15:14:11 \\
 Electrical mast & $24\times24\times72$ & $98.23$  & $107.53$& $0.100$  & 400& 01:14:20  \\
 Electrical mast & $48\times48\times144$ & $98.09$ & $104.28$& $0.100$  & 400& 07:16:38  \\
 L-beam  & $48\times48\times24$ & $590.56$ & $668.37$& $0.100$  & 400& 02:02:02  \\
 L-beam  & $96\times96\times48$ & $567.62$& $608.56$  & $0.100$  & 400&  13:32:11 \\
\hline
 \end{tabular}
\end{table}

It can be seen that even the example with the finest discretization (using approximately 1.2 million degrees of freedom) does not require more than 16 CPU hours on the workstation, making the computational cost manageable. Furthermore, it should be mentioned that there is great potential for a speed up if the code is run in parallel or in a lower-level programming language (e.g. C++) compared to MATLAB. Furthermore, it is interesting to note that in most parts of the design domain, only one or two layers have a finite width. This observation is in agreement with the work of~\citet{Bib:GibianskyCherkaev1987}, where it is shown that not all loading situations result in laminates of third rank. Finally, reference compliance values $\mathcal{J}^{ref}$ for designs obtained using the well-known Solid Isotropic Microstructure with Penalty (SIMP) method obtained on the same mesh sizes are shown in Table~\ref{Tab:ExpHbased.1} as well. These reference values have been obtained using a continuation scheme on the penalization factor $(p)$. We start with $p=1$ and slowly increase this to $p=4$ such that the design converges within 450 iterations. Furthermore, it should be noted that the density filter is used in these examples with a filter radius of $R = 2h^{c}$, where the filter is turned off in the final iterations to allow for black and white designs. Furthermore, it has to be noted that the void material is modeled using a Young's modulus $E^{-}=10^{-9}E^{+}$.

As expected the homogenization-based designs perform much better than the single-scale density-based designs. The reason is obvious, for the multi-scale designs optimal rank-3 microstructures are used, which contain much more information on this coarse mesh than the isotropic material using the SIMP method. Only for Michell's torsion sphere the values are close, this is expected since it is well-known that the optimal solution is a closed sphere of solid material~\citep{Bib:SigmundMichell}.

\subsection{De-homogenization}
As discussed before, the de-homogenization consist of two parts. The first part is the sorting of the vector fields. The second part is calculating mapping functions $\phi_{i}$, applying an average unit-cell spacing $\varepsilon$ and a length-scale $f_{min}$. The de-homogenization is performed in a multi-scale fashion, where the vector fields are sorted on coarse optimization mesh $\mathcal{T}^{c}$, the mapping functions are calculated on $\mathcal{T}^{i}$, while the design is projected on fine mesh $\mathcal{T}^{f}$. The total computational cost to de-homogenize each of the above-mentioned examples to a fine mesh of approximately 200 million elements is shown in Table~\ref{Tab:DeHomo.1}. For each example an average unit-cell spacing $\varepsilon = 40~h^{f}$ is chosen, with $f_{min} = 2~h^{f}$; however, it is noted that changing these values does not affect the computational cost. 

\begin{table}[ht!]
\centering
\caption{Different mesh sizes $\mathcal{T}^{c}$, and $\mathcal{T}^{i}$, $\mathcal{T}^{f}$, and computational cost for the de-homogenization step $T^{\phi}$.}
\label{Tab:DeHomo.1}
\centering
\setlength{\arrayrulewidth}{1pt} 
\begin{tabular}{ccccc}
 \hline
 \textbf{Example} & $\mathcal{T}^{c}$ & $\mathcal{T}^{i}$ & $\mathcal{T}^{f}$ & $T^{\phi}~[hh:mm:ss]$ \\ \hline
 Michell cantilever & $48\times24\times24$ & $192\times96\times96$ & $960\times480\times480$&  00:15:36 \\
 Michell cantilever & $96\times48\times48$ & $192\times96\times96$ & $960\times480\times480$&  00:45:39 \\
 Michell's torsion sphere & $48\times48\times48$ & $144\times144\times144$ & $576\times576\times576$& 00:36:28  \\
 Michell's torsion sphere & $72\times72\times72$ & $144\times144\times144$ & $576\times576\times576$&  01:19:44 \\
 Electrical mast & $24\times24\times72$ & $96\times96\times288$ & $384\times384\times1152 $& 00:21:43 \\
 Electrical mast & $48\times48\times144$ & $96\times96\times288$ & $384\times384\times1152 $& 00:54:49 \\
 L-beam  & $48\times48\times24$ & $192\time192\times96$ & $768\times768\times384 $&  00:23:32\\
 L-beam  & $96\times96\times48$ & $192\time192\times96$ & $768\times768\times384 $&  01:08:06 \\
\hline
 \end{tabular}
\end{table}

As can be seen all examples can be de-homogenized into very fine designs in less than one and a half hours, rendering the total computational cost for obtaining very fine designs in the range of approximately $2-17$ hours, utilizing only a single core on the workstation PC.

\subsection{Performance of the de-homogenized designs}
The next step is to demonstrate the performance of the presented approach. First, we show the effect of changing minimum length-scale $f_{min}$ and average unit-cell spacing $\varepsilon$ on the volume fraction of the de-homogenized design $V_{f}^{\phi}$. To do so, we use the Electrical mast example projected on a fine mesh of $384\times384\times1152$ elements. The results for both optimization meshes $\mathcal{T}^{c}$ are shown in Table~\ref{Tab:DeHomo.2}. 
\begin{table}[ht!]
\centering
\caption{The volume fraction $V_{f}^{\phi}$, compliance $\mathcal{J}^{\phi}$, stiffness per weight measure $\mathcal{S}^{\phi}$ and total computation time $T^{tot}$ of the de-homogenized electrical mast on a fine mesh of $384\times384\times1152$ elements. Furthermore, the volume fraction $V_{f}^{post}$ and compliance $\mathcal{J}^{post}$ and stiffness per weight  $\mathcal{S}^{post}$ after the post-processing step are shown. Results are shown for different $\mathcal{T}^{c}$, $f_{min}$ and $\varepsilon$.}
\label{Tab:DeHomo.2}
\centering
\setlength{\arrayrulewidth}{1pt} 
\begin{tabular}{ccccccc|ccc}
 \hline
  $\mathcal{T}^{c}$ & $\varepsilon$ & $f_{min}$ & $V_{f}^{\phi}$ & $\mathcal{J}^{\phi}$ & $\mathcal{S}^{\phi}$ &$T^{tot}~[hh:mm:ss]$ & $V_{f}^{post}$ & $\mathcal{J}^{post}$& $\mathcal{S}^{post}$\\ \hline
  $24\times24\times72$ & $24~h^{f}$ & $0~h^{f}$ & 0.0993 & 106.17 & 10.544& 01:36:03 & 0.0992 & 106.17 &10.528 \\
  $24\times24\times72$ & $24~h^{f}$ & $2~h^{f}$ & 0.1046 & 101.83 & 10.650 & & 0.1041& 101.83 &10.598\\
  $24\times24\times72$ & $24~h^{f}$ & $3~h^{f}$ & 0.1093 & 98.96 & 10.814& & 0.1086& 98.42 &10.692\\
  $24\times24\times72$ & $24~h^{f}$ & $4~h^{f}$ & 0.1156 & 95.19 & 11.003& &  0.1151& 93.83&10.796 \\
  $24\times24\times72$ & $32~h^{f}$ & $0~h^{f}$ & 0.0990 & 107.87 & 10.678 & & 0.0987 & 107.86&10.650 \\
  $24\times24\times72$ & $32~h^{f}$ & $2~h^{f}$ & 0.1020 & 105.08 & 10.720& & 0.1015& 105.06 &10.669\\
  $24\times24\times72$ & $32~h^{f}$ & $3~h^{f}$ & 0.1050 & 102.81 & 10.799& & 0.1043  & 102.40&10.680 \\
  $24\times24\times72$ & $32~h^{f}$ & $4~h^{f}$ & 0.1088 & 100.42 & 10.923& & 0.1077 & 99.44 &10.713\\
  $24\times24\times72$ & $40~h^{f}$ & $0~h^{f}$ & 0.0979 & 110.80 & 10.843& & 0.0976& 110.78 &10.812\\
  $24\times24\times72$ & $40~h^{f}$ & $2~h^{f}$ & 0.0996 & 109.11 & 10.870& & 0.0993 & 109.11&10.837 \\
  $24\times24\times72$ & $40~h^{f}$ & $3~h^{f}$ & 0.1016 & 107.64 & 10.933& & 0.1008& 107.36&10.824\\
  $24\times24\times72$ & $40~h^{f}$ & $4~h^{f}$ & 0.1040 & 105.78 & 10.996& & 0.1028 & 105.27 &10.822\\ 
\hline
  $48\times48\times144$ & $24~h^{f}$ & $0~h^{f}$ & 0.0994 & 108.14  &10.755 & 08:11:27 & 0.0993 & 108.14 &10.737\\
  $48\times48\times144$& $24~h^{f}$ & $2~h^{f}$ & 0.1059& 102.52&10.858  & & 0.1054 & 102.50 &10.805\\
  $48\times48\times144$& $24~h^{f}$ & $3~h^{f}$ & 0.1121& 98.21 &11.008& & 0.1115 & 97.37 &10.853\\
  $48\times48\times144$& $24~h^{f}$ & $4~h^{f}$ & 0.1195 & 93.64 &11.194& & 0.1198 & 91.70 &10.984\\
  $48\times48\times144$& $32~h^{f}$ & $0~h^{f}$ & 0.0974&  114.03 &11.109& &0.0972& 114.03 &11.080\\
  $48\times48\times144$& $32~h^{f}$ & $2~h^{f}$ & 0.1013&  109.98 &11.145& & 0.1007& 109.99 &11.083\\
  $48\times48\times144$& $32~h^{f}$ & $3~h^{f}$ & 0.1054 & 106.69 &11.240& & 0.1042 & 106.38 &11.088\\
  $48\times48\times144$& $32~h^{f}$ & $4~h^{f}$ & 0.1105 & 102.75 &11.350& & 0.1088& 102.00 &11.097\\
  $48\times48\times144$& $40~h^{f}$ & $0~h^{f}$ & 0.0979 & 113.90 &11.153& & 0.0976& 113.90 &11.119\\
  $48\times48\times144$& $40~h^{f}$ & $2~h^{f}$ & 0.1006 & 111.26 &11.189& & 0.1001 & 111.26 &11.139\\
  $48\times48\times144$& $40~h^{f}$ & $3~h^{f}$ & 0.1035 & 108.89 &11.272& & 0.1026& 108.15 &11.098\\
  $48\times48\times144$& $40~h^{f}$ & $4~h^{f}$ & 0.1071 & 106.08 &11.364& & 0.1058& 104.76 &11.086\\ 
\hline
 \end{tabular}
\end{table}

 As can be seen, adding a minimum length-scale adds a significant amount of material, hence the volume constraint can be violated up to $0.20~V_{max}$. Nevertheless, this effect can be minimized if reasonable values for $\varepsilon$ and $f_{min}$ are chosen. When $f_{min}$ is large compared to $\varepsilon$, a large amount of material will be added to satisfy the minimum length-scale, which results in a large violation of the volume constraint. Generally, a good rule of thumb is that $\frac{\varepsilon}{f_{min}} > 10$, such that the violation of the volume constraint stays below $10\%$ of $V_{max}$. Furthermore, it is obvious that $\varepsilon$ has to be small compared to the domain size, otherwise the de-homogenized design cannot represent the homogenization-based design. Ideally, we would thus like to de-homogenize the multi-scale design to a finer mesh, see $e.g.$~\citet{Bib:GroenSigmund2017}, since this would allow finer details and smaller or no violations of the volume constraint, while a minimum length-scale can still be guaranteed, ideally related to $\eta$ used in the homogenization-based topology optimization procedure.

 Furthermore, all the de-homogenized designs perform close to the homogenization-based solutions. This can be verified by analyzing the compliance of the de-homogenized designs on the fine mesh $\mathcal{J}^{\phi}$, which is also shown in Table~\ref{Tab:DeHomo.2}. As expected the designs which have a higher volume have a lower compliance. Nevertheless, it can be seen that all values are close to the homogenization-based compliance $\mathcal{J}^{c}$. It can be seen that a smaller $\varepsilon$ results in a better performance, which is expected, since the de-homogenized designs better resemble the multi-scale designs. Hence, this again shows the need for an even finer resolution of the de-homogenized designs. 
 Furthermore, we introduce an additional measure $\mathcal{S}^{\phi}$, which is the compliance multiplied by the volume and can be seen as a measure of stiffness per volume. The lower this value, the better the material usage and hence performance. The use of $\mathcal{S}^{\phi}$ allows us to quantitatively compare the different de-homogenized designs with each other to show that an almost constant performance is achieved for the different values of $\varepsilon$ and $f_{min}$. However, with a slight advantage for the designs without length-scale enforcement. The values for the stiffness per volume of the homogenization-based structures are $\mathcal{S}^{c}=9.823$ and $\mathcal{S}^{c}=9.809$ for the coarse and slightly finer mesh respectively.

\begin{figure}[h!]
\centering
\subfloat[$\varepsilon = 24~h^{f}$, $f_{min}=0~h^{f}$.]{\includegraphics[width=0.25\textwidth]{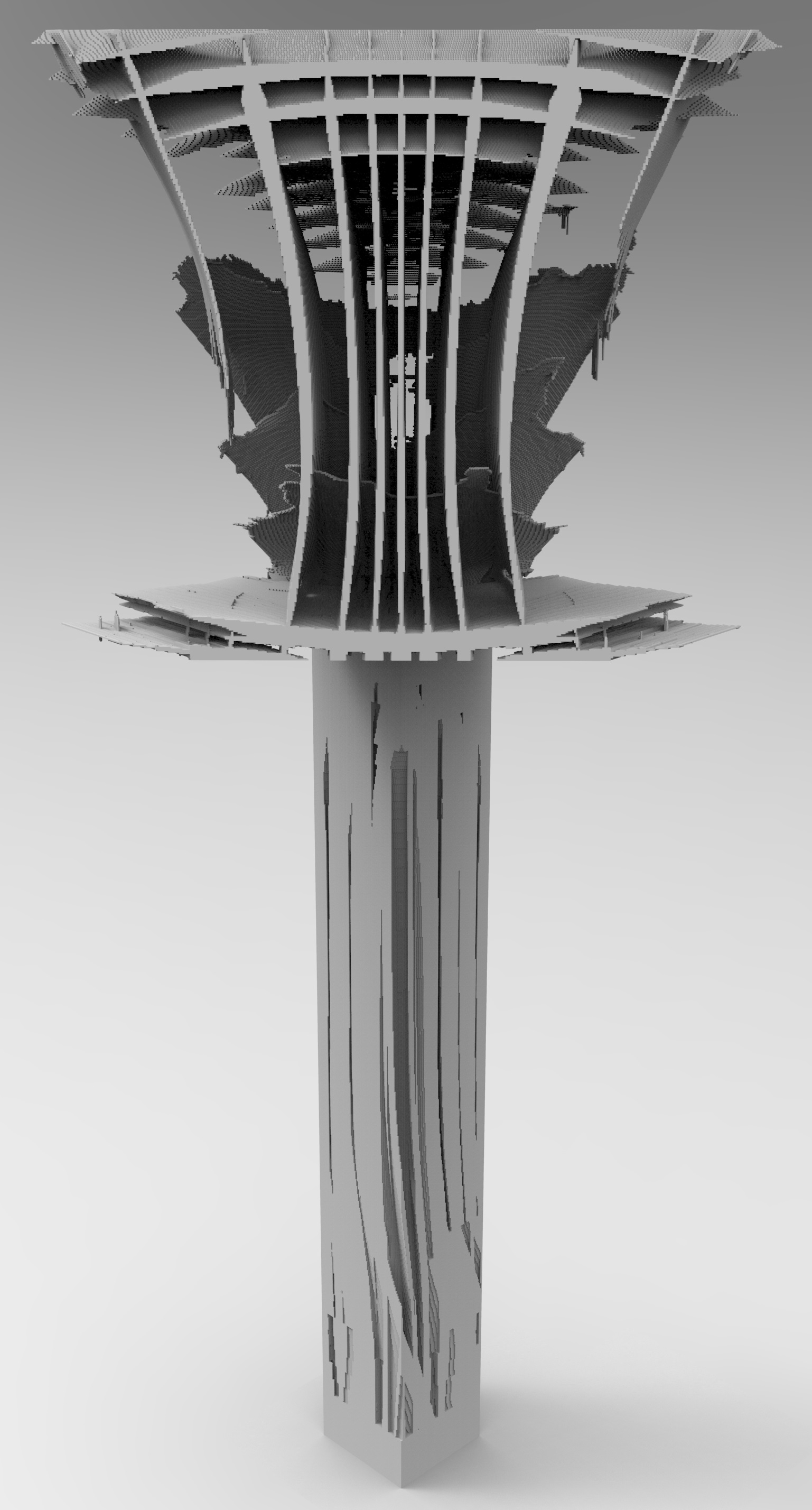}} 
\subfloat[$\varepsilon = 40~h^{f}$, $f_{min}=2~h^{f}$.]{\includegraphics[width=0.25\textwidth]{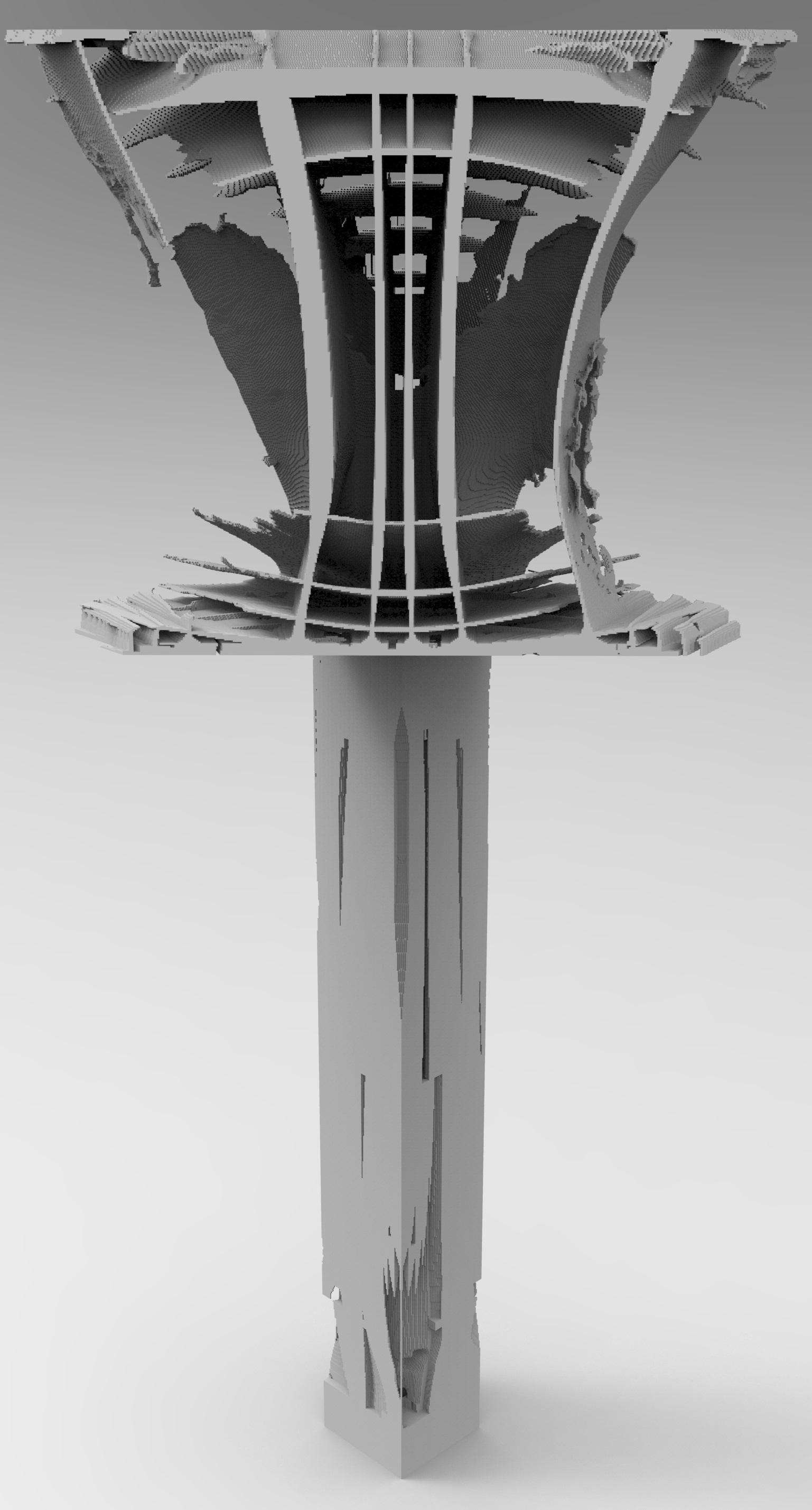}} 
\subfloat[$\varepsilon = 24~h^{f}$, $f_{min}=3~h^{f}$.]{\includegraphics[width=0.25\textwidth]{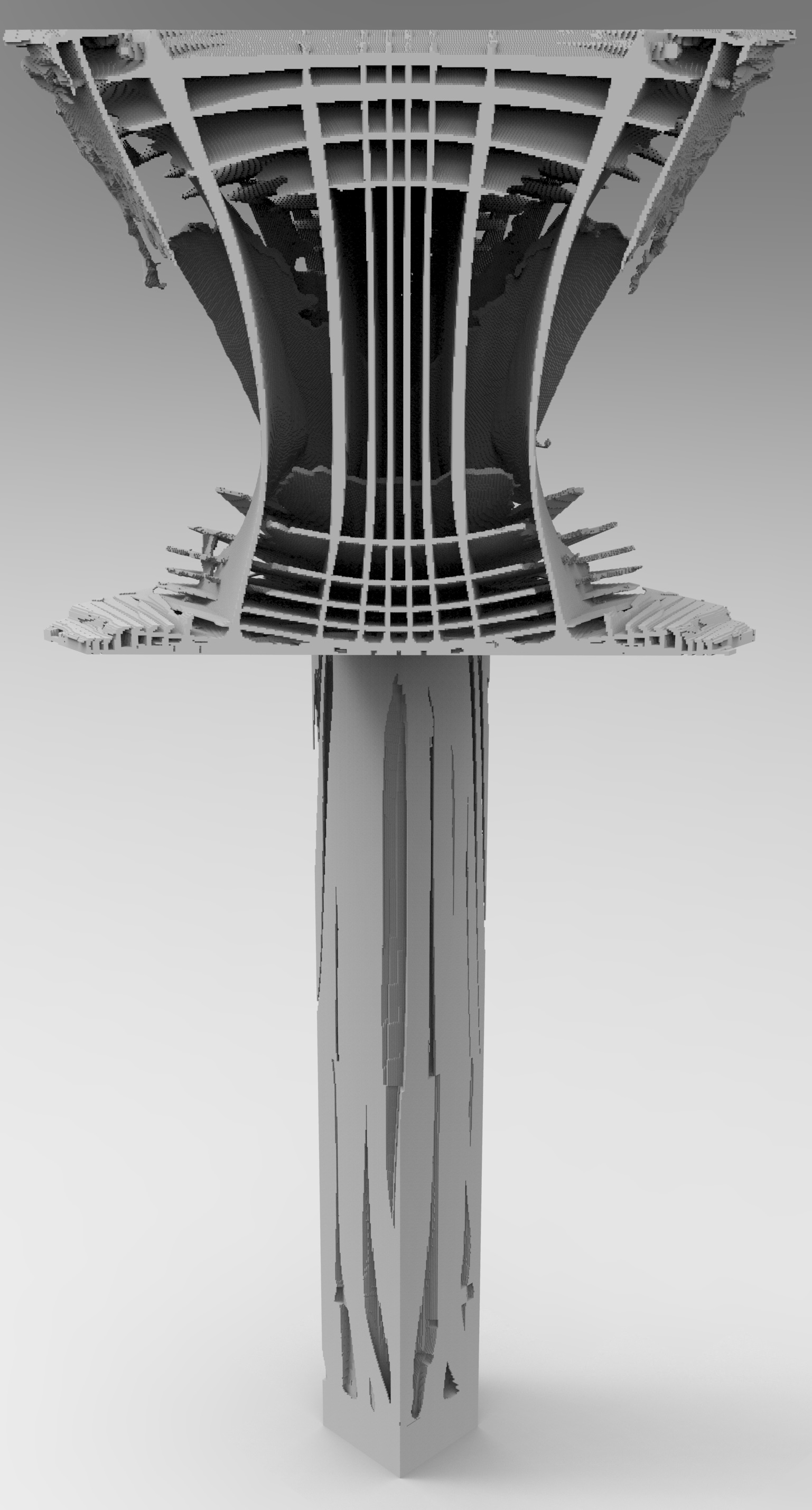}} 
\subfloat[$\varepsilon = 24~h^{f}$, $f_{min}=3~h^{f}$, after post-processing.]{\includegraphics[width=0.25\textwidth]{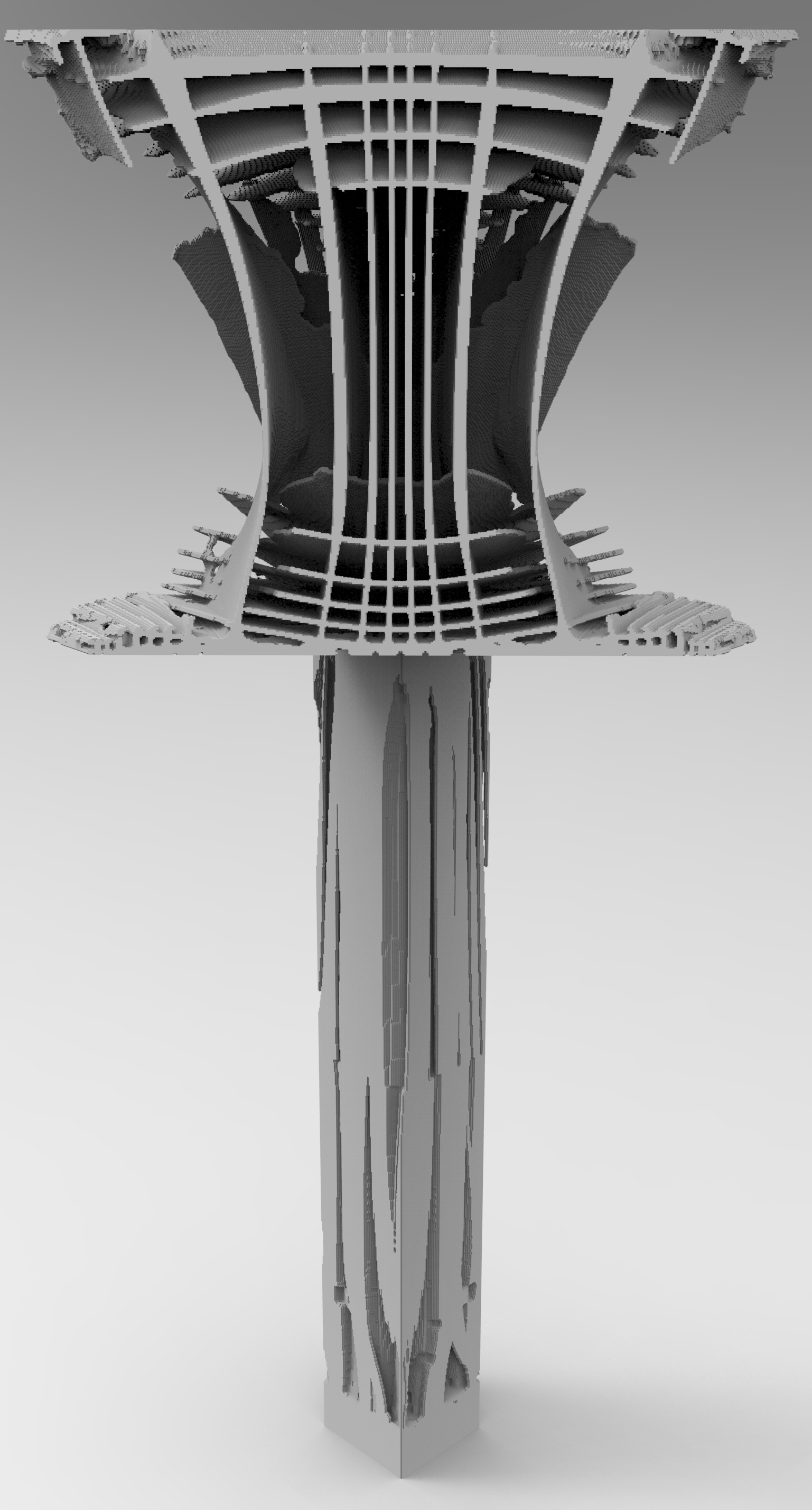}} 
\caption{Cut-outs of the de-homogenized designs of the electrical mast example de-homogenized on a fine mesh of $1152\times384\times384$ voxels for different values of $\varepsilon$ and $f_{min}$. Example (a) is optimized on $\mathcal{T}^{c} =24\times24\times72$, while the other examples are obtained using $\mathcal{T}^{c} =48\times48\times144$. For other views of the design in Figure~\ref{Fig:DeHomo.1}(d), see Figures~\ref{Fig:DeHomo.2}(a) and (b).}
\label{Fig:DeHomo.1}
\end{figure}

It is also interesting to see that the mesh on which the homogenization-based designs are achieved can be very coarse. This demonstrates how much information is captured by using a rank-3 parameterization. Furthermore, this shows that the total time $T^{tot}$ to obtain the de-homogenized designs without evaluating the fine-scale compliance can be as low as one and a half hour on a single PC. The times are only shown once per optimization mesh, since the choice of $\varepsilon$ and $f_{min}$ does not affect the calculation of mapping functions $\phi_{i}$. Section views of the de-homogenized designs for different values of $\varepsilon$ and $f_{min}$ are shown in Figures~\ref{Fig:DeHomo.1}(a),(b) and (c). 

\begin{figure}[h!]
\centering
\subfloat[De-homogenized.]{\includegraphics[width=0.2135\textwidth]{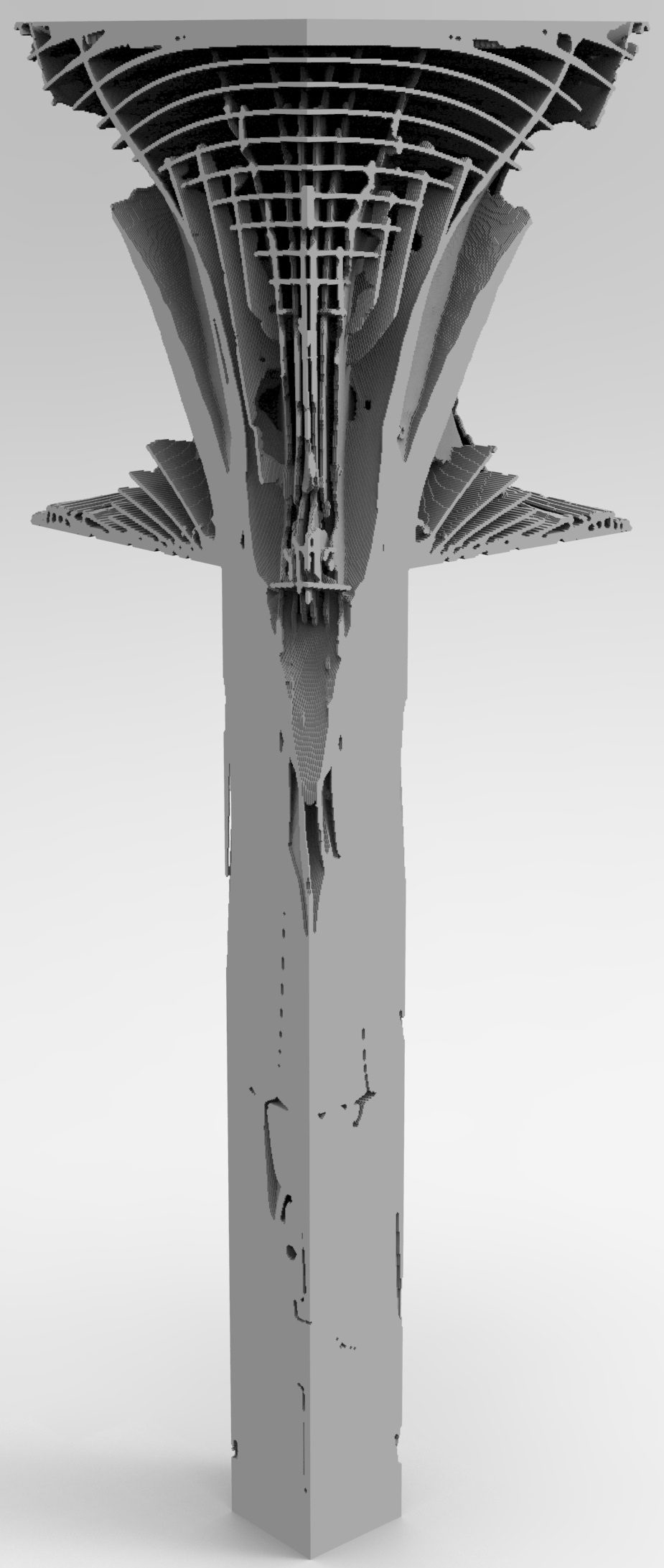}} 
\subfloat[De-homogenized.]{\includegraphics[width=0.2865\textwidth]{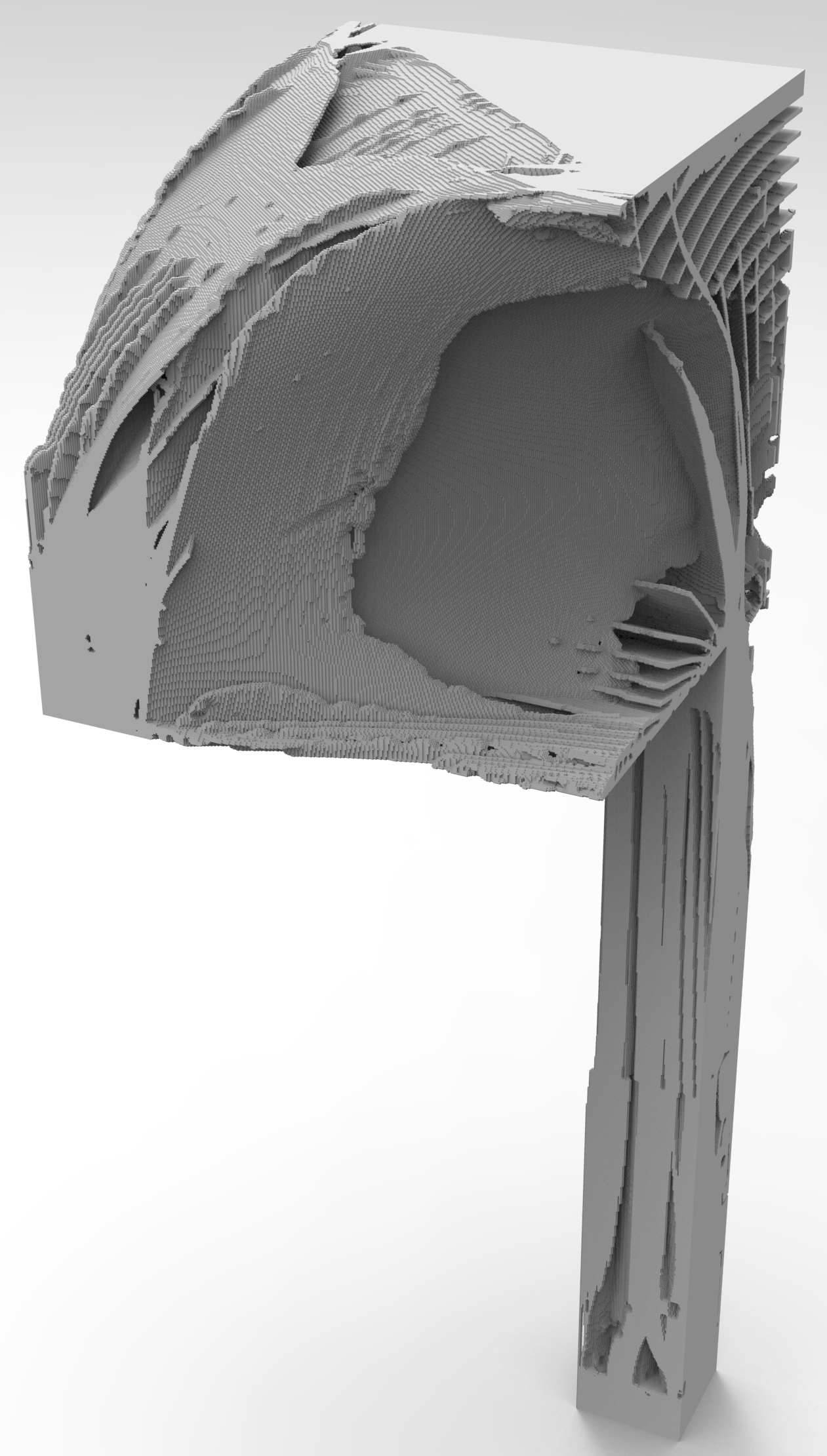}} 
\subfloat[Density-based.]{\includegraphics[width=0.2135\textwidth]{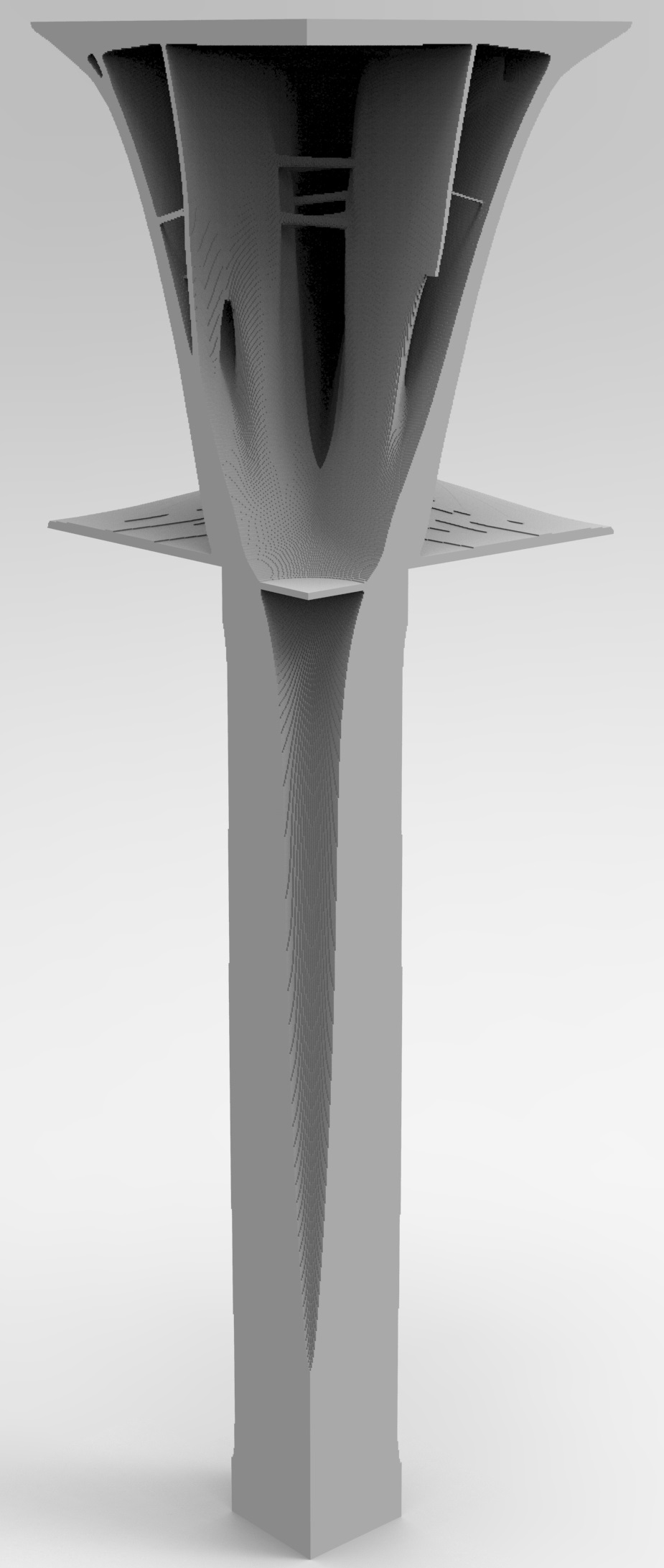}} 
\subfloat[Density-based.]{\includegraphics[width=0.2865\textwidth]{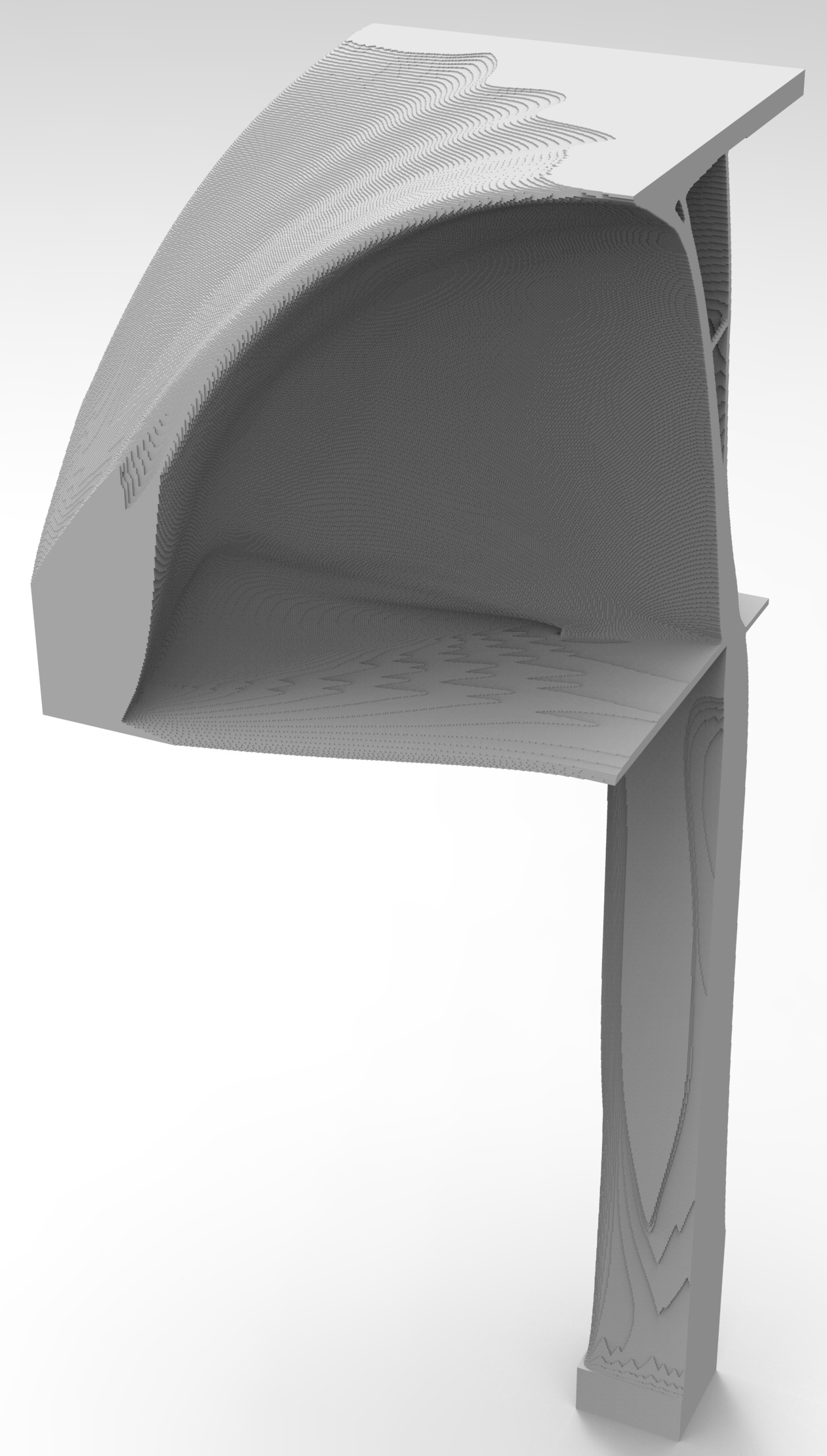}} 
\caption{The de-homogenized design of the electrical mast on a fine mesh of $1152\times384\times384$ voxels for $\varepsilon = 24~h^{f}$ and $f_{min}=3~h^{f}$, after post-processing. Furthermore, a reference design using density-based topology optimization is shown.}
\label{Fig:DeHomo.2}
\end{figure}

From these Figures it can also be seen that the average unit-cell spacing $\varepsilon$ has to be small enough to allow for the de-homogenized design to represent the homogenization-based design. Furthermore, it can be seen that the de-homogenized design obtained using $\mathcal{T}^{c} =48\times48\times144$ contains more microstructural details. Finally, the post-processing method described in the previous section can be used to get rid of some members that do not carry any load. An example of a de-homogenized design including this post-processing step can be seen in Figure~\ref{Fig:DeHomo.1}(d) and Figures~\ref{Fig:DeHomo.2}(a) and (b). The corresponding values for the volume fraction $V_{f}^{post}$, compliance $\mathcal{J}^{post}$ and measure for stiffness per weight $\mathcal{S}^{post}$ can be seen in Table~\ref{Tab:DeHomo.2} as well. Here it should be noted that in general the post-processing step not only removes material but also improves the stiffness per weight measure. Unfortunately however, this requires several fine-scale analyses that come at a large computational cost, $i.e.$ using 3200 cores on the DTU Sophia cluster.

The performance of the de-homogenized designs of the Michell cantilever, Michell's torsion sphere and the L-shaped beam are shown in Tables~\ref{Tab:DeHomo.3}-\ref{Tab:DeHomo.5}. The values for the stiffness per volume of the homogenization-based structures are $\mathcal{S}^{c}=22.79$ and $\mathcal{S}^{c}=22.67$ for the coarse and slightly finer mesh respectively for the Michell cantilever, $\mathcal{S}^{c}=1.233$ and $\mathcal{S}^{c}=1.400$ for Michell's torsion sphere and $\mathcal{S}^{c}=59.06$ and $\mathcal{S}^{c}=56.76$ for the L-shaped beam.

\begin{table}[ht!]
\centering
\caption{The volume fraction $V_{f}^{\phi}$, compliance $\mathcal{J}^{\phi}$, stiffness per weight measure $\mathcal{S}^{\phi}$ and total computation time $T^{tot}$ of the de-homogenized Michell cantilever on a fine mesh of $960\times960\times480$ elements. Furthermore, the volume fraction $V_{f}^{post}$ and compliance $\mathcal{J}^{post}$ and stiffness per weight  $\mathcal{S}^{post}$ after the post-processing step are shown. Results are shown for different $\mathcal{T}^{c}$, $f_{min}$ and $\varepsilon$.}
\label{Tab:DeHomo.3}
\centering
\setlength{\arrayrulewidth}{1pt} 
\begin{tabular}{ccccccc|ccc}
 \hline
  $\mathcal{T}^{c}$ & $\varepsilon$ & $f_{min}$ & $V_{f}^{\phi}$ & $\mathcal{J}^{\phi}$ & $\mathcal{S}^{\phi}$ &$T^{tot}~[hh:mm:ss]$ & $V_{f}^{post}$ & $\mathcal{J}^{post}$& $\mathcal{S}^{post}$\\ \hline
  $48\times24\times24$& $24~h^{f}$ & $0~h^{f}$ & 0.1004 & 253.37 & 25.427 & 01:39:27 & 0.0993 & 253.39 & 25.171 \\
  $48\times24\times24$& $24~h^{f}$ & $2~h^{f}$ & 0.1109 & 232.91 & 25.825 & & 0.1099 & 232.92 & 25.591 \\
  $48\times24\times24$& $24~h^{f}$ & $3~h^{f}$ & 0.1228 & 214.72 & 26.367&  & 0.1202 & 215.45 & 25.888\\
  $48\times24\times24$& $24~h^{f}$ & $4~h^{f}$ & 0.1378 & 195.72 & 26.978&  & 0.1343 & 195.50 &26.258 \\
  $48\times24\times24$& $32~h^{f}$ & $0~h^{f}$ & 0.1018 & 245.15 & 24.947 & & 0.1012 & 245.15 & 24.801 \\
  $48\times24\times24$& $32~h^{f}$ & $2~h^{f}$ & 0.1081 & 234.41 & 25.345& & 0.1076 & 234.56 & 25.241 \\
  $48\times24\times24$& $32~h^{f}$ & $3~h^{f}$ & 0.1153 & 222.83 & 25.697& & 0.1137 & 1152.2 & 130.98\\
  $48\times24\times24$& $32~h^{f}$ & $4~h^{f}$ & 0.1250 & 209.35 & 26.171 &   & 0.1229 & 209.79 & 25.784 \\
  $48\times24\times24$& $40~h^{f}$ & $0~h^{f}$ & 0.1017 & 247.14 & 25.143 &  & 0.1010 & 247.15 & 24.952 \\
  $48\times24\times24$& $40~h^{f}$ & $2~h^{f}$ & 0.1058 & 240.84 & 25.480 & & 0.1051 & 240.86 & 25.321 \\
  $48\times24\times24$& $40~h^{f}$ & $3~h^{f}$ & 0.1102 & 234.62 & 25.848 &  & 0.1082 & 235,31 & 25.449 \\
  $48\times24\times24$& $40~h^{f}$ & $4~h^{f}$ & 0.1165 & 224.94 &26.214 &  & 0.1140 & 225.64 &25.712 \\ 
\hline
  $96\times48\times48$ & $24~h^{f}$ & $0~h^{f}$ & 0.1008 & 251.43 & 25.344& 10:31:14 & 0.1004 & 251.43 & 25.231 \\
  $96\times48\times48$ & $24~h^{f}$ & $2~h^{f}$ & 0.1073 & 235.96 & 25.322 & & 0.1069 & 236.55 & 25.281 \\
  $96\times48\times48$ & $24~h^{f}$ & $3~h^{f}$ & 0.1162 & 220.87 & 25.670 &  & 0.1149 & 222.73 & 25.598 \\
  $96\times48\times48$ & $24~h^{f}$ & $4~h^{f}$ & 0.1306 & 200.48 & 26.184 &  & 0.1287 & 201.98 & 25.988 \\
  $96\times48\times48$ & $32~h^{f}$ & $0~h^{f}$ & 0.1024 & 245.40 & 25.136  &  & 0.1019 & 245.42 & 25.010\\
  $96\times48\times48$ & $32~h^{f}$ & $2~h^{f}$ & 0.1061 & 237.64 & 25.219 &  & 0.1055 & 237.86 & 25.104\\
  $96\times48\times48$ & $32~h^{f}$ & $3~h^{f}$ & 0.1105 & 230.28 & 25.437&  & 0.1089 & 231.80 & 25.236\\
  $96\times48\times48$ & $32~h^{f}$ & $4~h^{f}$ & 0.1179 & 218.49 &25.759 &  & 0.1156 & 220.10 & 25.447\\
  $96\times48\times48$ & $40~h^{f}$ & $0~h^{f}$ & 0.1026 & 243.29 & 24.966 &  & 0.1021 & 243.31 & 24.845\\
  $96\times48\times48$ & $40~h^{f}$ & $2~h^{f}$ & 0.1052 & 237.14 & 24.942 &  & 0.1049 & 238.70 & 25.034 \\
  $96\times48\times48$ & $40~h^{f}$ & $3~h^{f}$ & 0.1087 & 232.77 & 25.292 &  & 0.1076 & 234.29 & 25.202\\
  $96\times48\times48$ & $40~h^{f}$ & $4~h^{f}$ & 0.1131 & 227.19 &25.706 & & 0.1117 & 228.36 &25.508\\ 
\hline
 \end{tabular}
\end{table}

\begin{figure}[h!]
\centering
\subfloat[De-homogenized design.]{\includegraphics[width=0.330\textwidth]{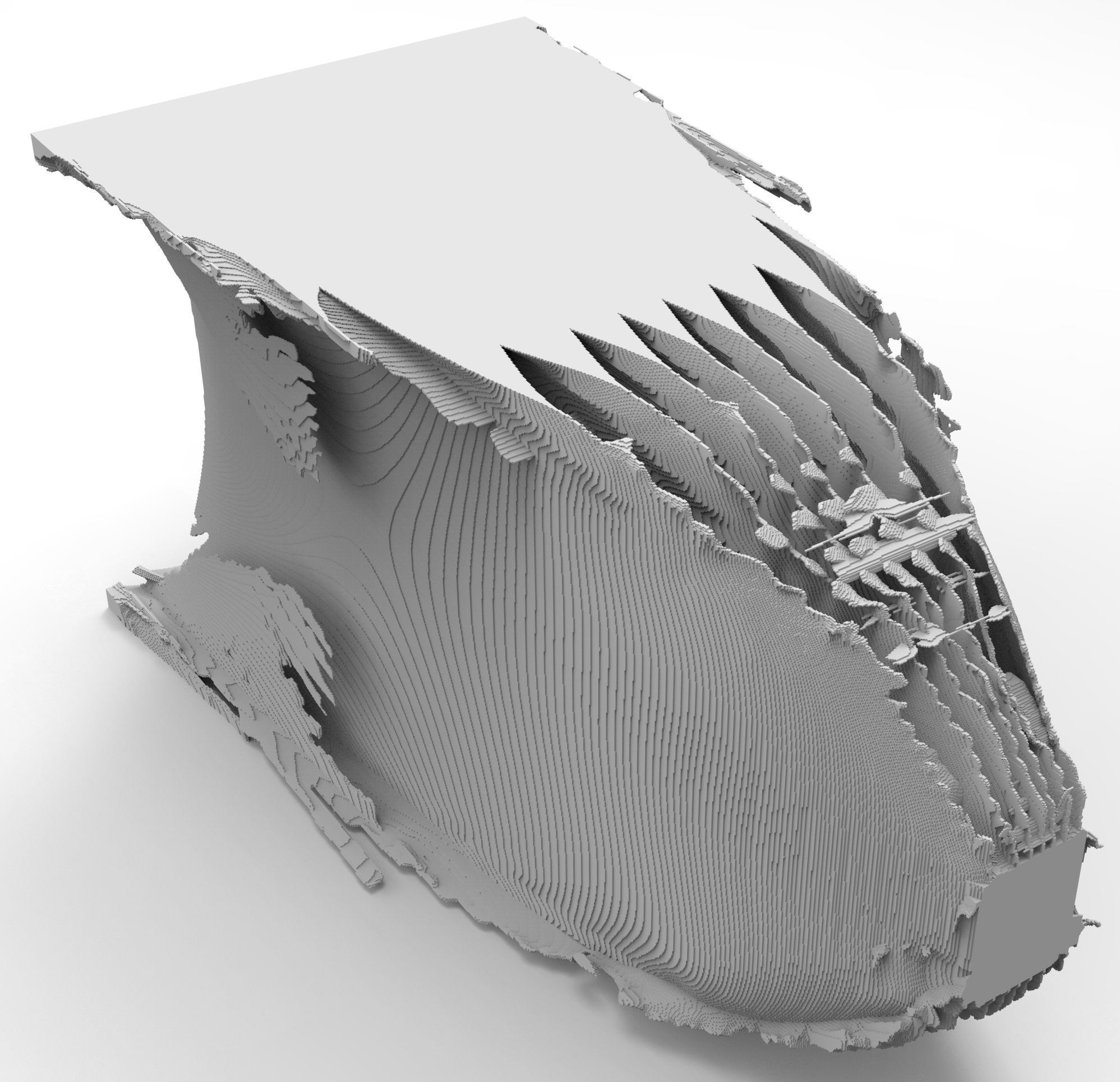}} 
\subfloat[After post-processing.]{\includegraphics[width=0.330\textwidth]{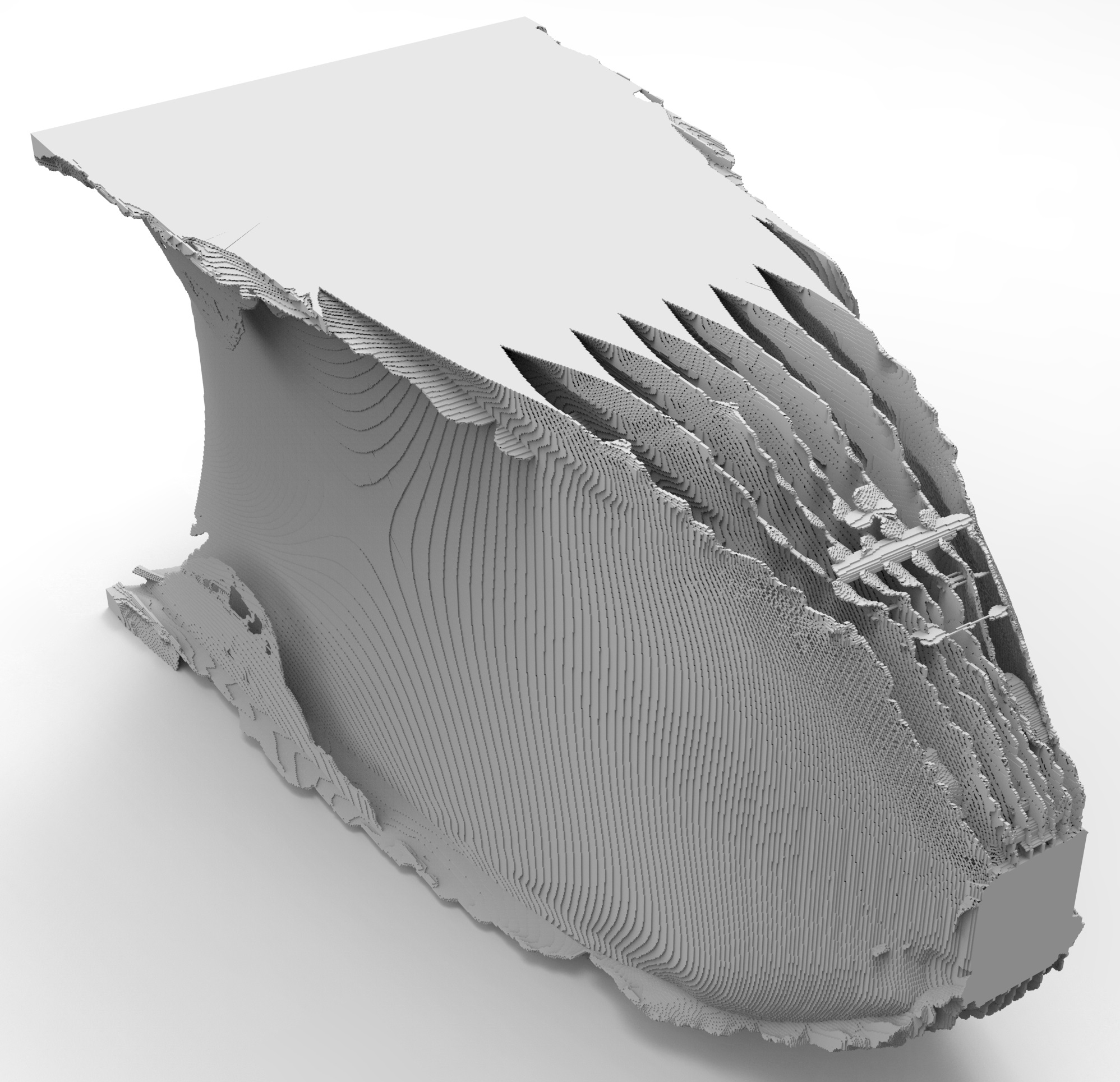}} 
\subfloat[After post-processing.]{\includegraphics[width=0.330\textwidth]{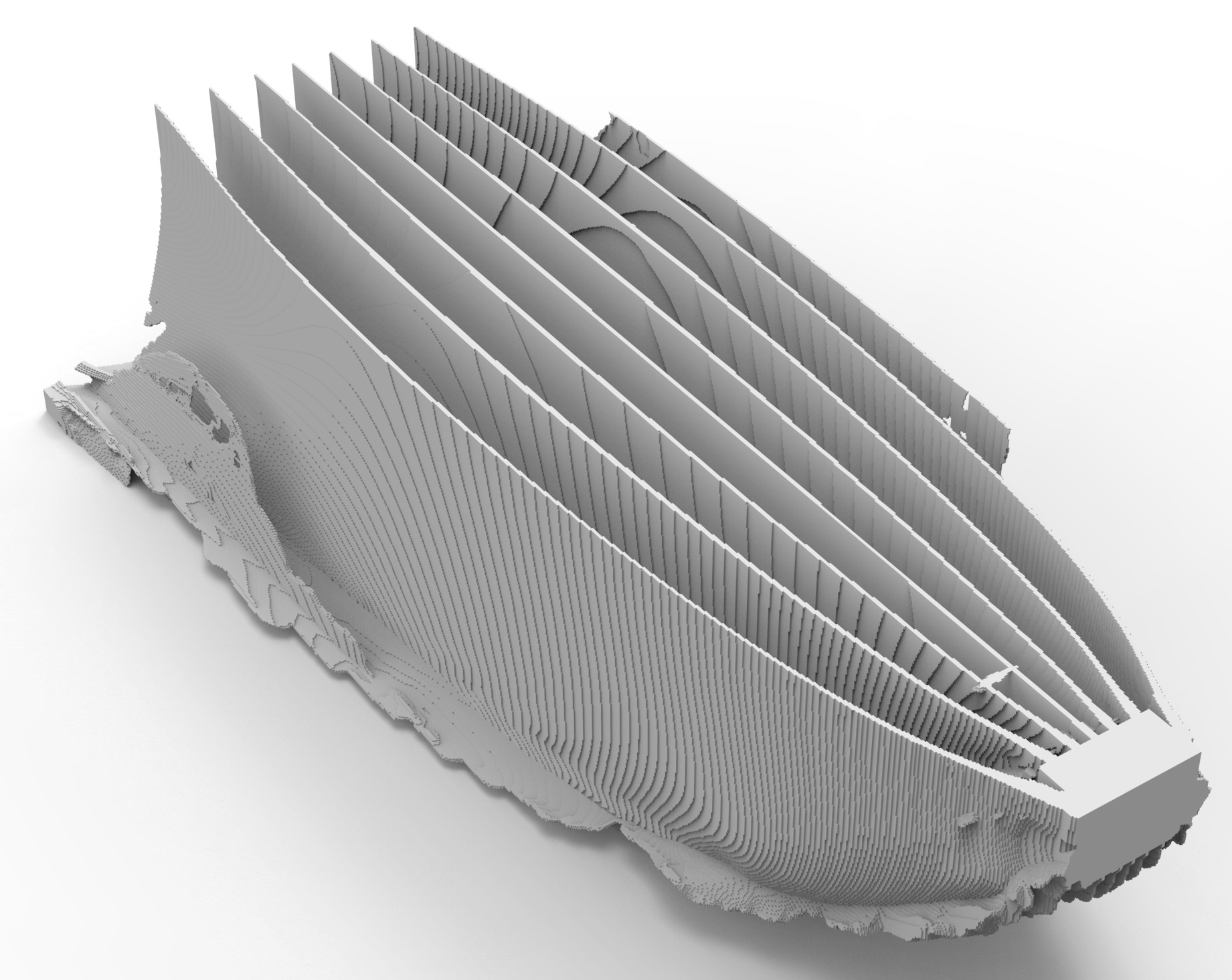}}
\caption{A Michell cantilever optimized on $\mathcal{T}^{c} = 96\times48\times48$ elements and de-homogenized on a fine mesh of $960\times480\times480$ using $\varepsilon=40~h^{f}$ and $f_{min} = 2~h^{f}$.}
\label{Fig:DeHomo.3}
\end{figure}

\begin{table}[ht!]
\centering
\caption{The volume fraction $V_{f}^{\phi}$, compliance $\mathcal{J}^{\phi}$, stiffness per weight measure $\mathcal{S}^{\phi}$ and total computation time $T^{tot}$ of Michell's torsion sphere de-homogenized on a fine mesh of $576\times576\times576$ elements. Furthermore, the volume fraction $V_{f}^{post}$ and compliance $\mathcal{J}^{post}$ and stiffness per weight  $\mathcal{S}^{post}$ after the post-processing step are shown. Results are shown for different $\mathcal{T}^{c}$, $f_{min}$ and $\varepsilon$.}
\label{Tab:DeHomo.4}
\centering
\setlength{\arrayrulewidth}{1pt} 
\begin{tabular}{ccccccc|ccc}
 \hline
  $\mathcal{T}^{c}$ & $\varepsilon$ & $f_{min}$ & $V_{f}^{\phi}$ & $\mathcal{J}^{\phi}$ & $\mathcal{S}^{\phi}$ &$T^{tot}~[hh:mm:ss]$ & $V_{f}^{post}$ & $\mathcal{J}^{post}$& $\mathcal{S}^{post}$\\ \hline
  $48\times48\times48$& $24~h^{f}$ & $0~h^{f}$ & 0.0955 & 20.77 & 1.983 & 05:05:03 & 0.0953 & 20.77 &1.979\\
  $48\times48\times48$& $24~h^{f}$ & $2~h^{f}$ & 0.0969 & 20.46 &1.981&  & 0.0959 & 20.55 & 1.971 \\
  $48\times48\times48$& $24~h^{f}$ & $3~h^{f}$ & 0.0988 & 20.53 & 2.029& & 0.0976 & 20.54 &2.005\\
  $48\times48\times48$& $24~h^{f}$ & $4~h^{f}$ & 0.1051 & 20.59 &2.164 & & 0.1040 & 20.57 &2.139 \\
  $48\times48\times48$& $32~h^{f}$ & $0~h^{f}$ & 0.0989 & 20.16 &1.995 & & 0.0854 & 20.87 &1.783\\
  $48\times48\times48$& $32~h^{f}$ & $2~h^{f}$ & 0.1023 & 20.44 &2.090 & & 0.1003 & 20.60 &2.065\\
  $48\times48\times48$& $32~h^{f}$ & $3~h^{f}$ & 0.1082 & 20.55 &2.225 & & 0.0886 & 20.75 &1.839\\
  $48\times48\times48$& $32~h^{f}$ & $4~h^{f}$ & 0.1152 & 19.82 &2.284&  & 0.0911 & 20.63 &1.879\\
  $48\times48\times48$& $40~h^{f}$ & $0~h^{f}$ & 0.0976 & 20.20 & 1.972& & 0.0943 & 20.66 &1.949\\
  $48\times48\times48$& $40~h^{f}$ & $2~h^{f}$ & 0.1012 & 19.84 & 2.007&  & 0.0976 & 20.72& 2.021 \\
  $48\times48\times48$& $40~h^{f}$ & $3~h^{f}$ & 0.1034 & 20.70 & 2.140& & 0.0947 & 20.73 &1.963\\
  $48\times48\times48$& $40~h^{f}$ & $4~h^{f}$ & 0.1055 & 20.58 &2.170&  & 0.0951 & 20.65 &1.964\\ 
\hline
  $72\times72\times72$ & $24~h^{f}$ & $0~h^{f}$ & 0.0968 & 20.71 &2.005 & 16:33:55 & 0.0953 & 20.71 & 1.974\\
  $72\times72\times72$ & $24~h^{f}$ & $2~h^{f}$ & 0.1018 & 20.60 &2.097&  & 0.1012 & 20.61 & 2.087\\
  $72\times72\times72$ & $24~h^{f}$ & $3~h^{f}$ & 0.1068 & 20.69 &2.210&  & 0.1062 & 20.69 &2.197\\
  $72\times72\times72$ & $24~h^{f}$ & $4~h^{f}$ & 0.1156 & 20.37 &2.355&  & 0.1128 & 20.58 &2.321\\
  $72\times72\times72$ & $32~h^{f}$ & $0~h^{f}$ & 0.0898 & 20.91 &1.878&  & 0.0826 & 20.93 &1.729\\
  $72\times72\times72$ & $32~h^{f}$ & $2~h^{f}$ & 0.0964 & 17.54 &1.690&  & 0.0833 & 20.84 &1.736\\
  $72\times72\times72$ & $32~h^{f}$ & $3~h^{f}$ & 0.1027 & 20.43 &2.098 &  & 0.0853 & 20.73 &1.768 \\
  $72\times72\times72$ & $32~h^{f}$ & $4~h^{f}$ & 0.1100 & 16.19 &1.781 &  & 0.0884 & 20.71 & 1.831 \\
  $72\times72\times72$ & $40~h^{f}$ & $0~h^{f}$ & 0.0984 & 20.70 &2.036 &  & 0.0983 & 20.70 & 2.034\\
  $72\times72\times72$ & $40~h^{f}$ & $2~h^{f}$ & 0.0992 & 20.57 &2.040 &  & 0.0988 & 20.58 &2.034\\
  $72\times72\times72$ & $40~h^{f}$ & $3~h^{f}$ & 0.1011 & 19.58 &1.980 &  & 0.1006 & 20.61&2.074 \\
  $72\times72\times72$ & $40~h^{f}$ & $4~h^{f}$ & 0.1035 & 20.67 &2.139 &  & 0.1030 & 20.67 &2.130\\ 
\hline
 \end{tabular}
\end{table}

\begin{figure}[h!]
\centering
\hfill
\subfloat[Full view.]{\includegraphics[width=0.4\textwidth]{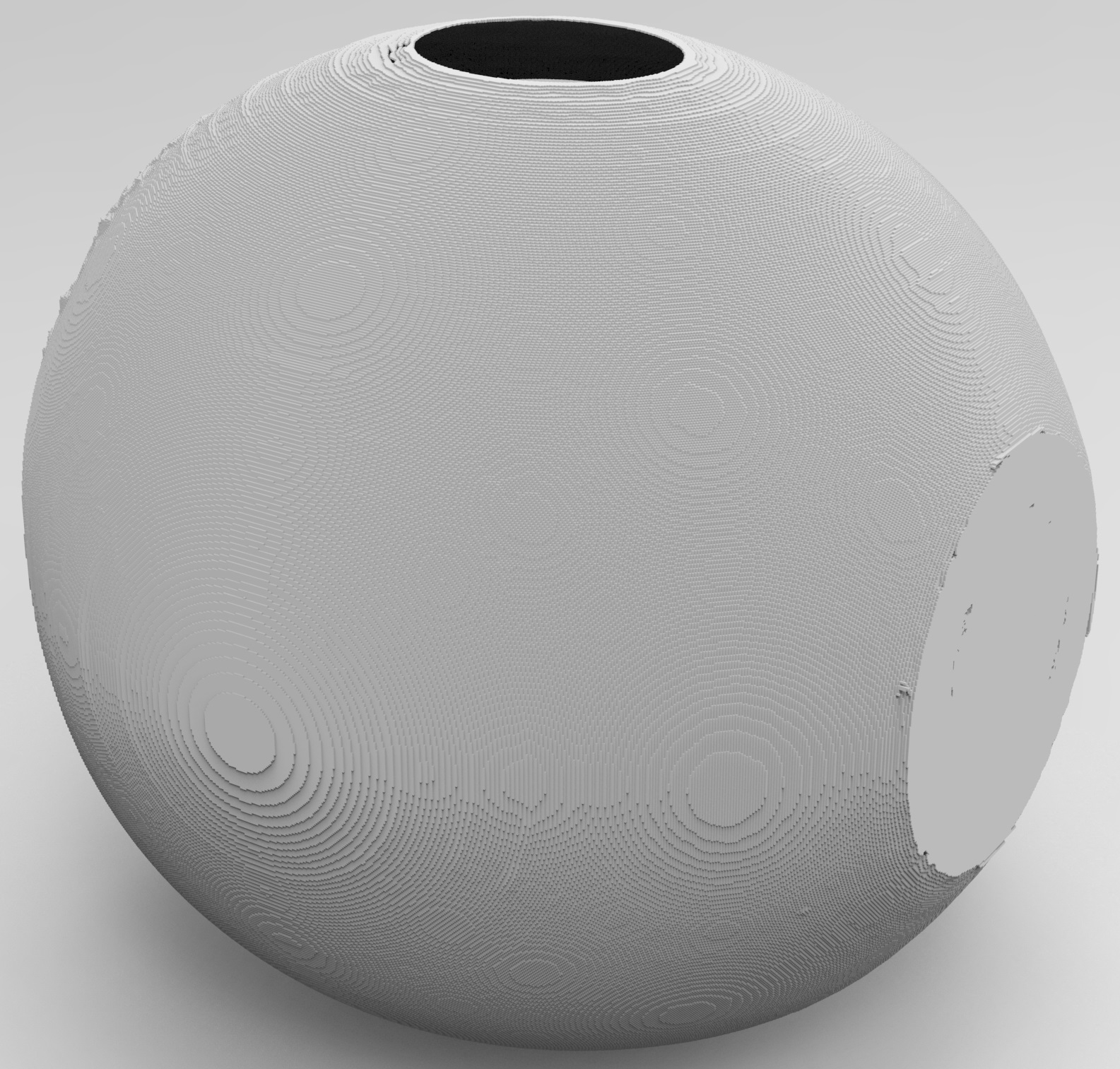}} \hfill
\subfloat[Section view.]{\includegraphics[width=0.36\textwidth]{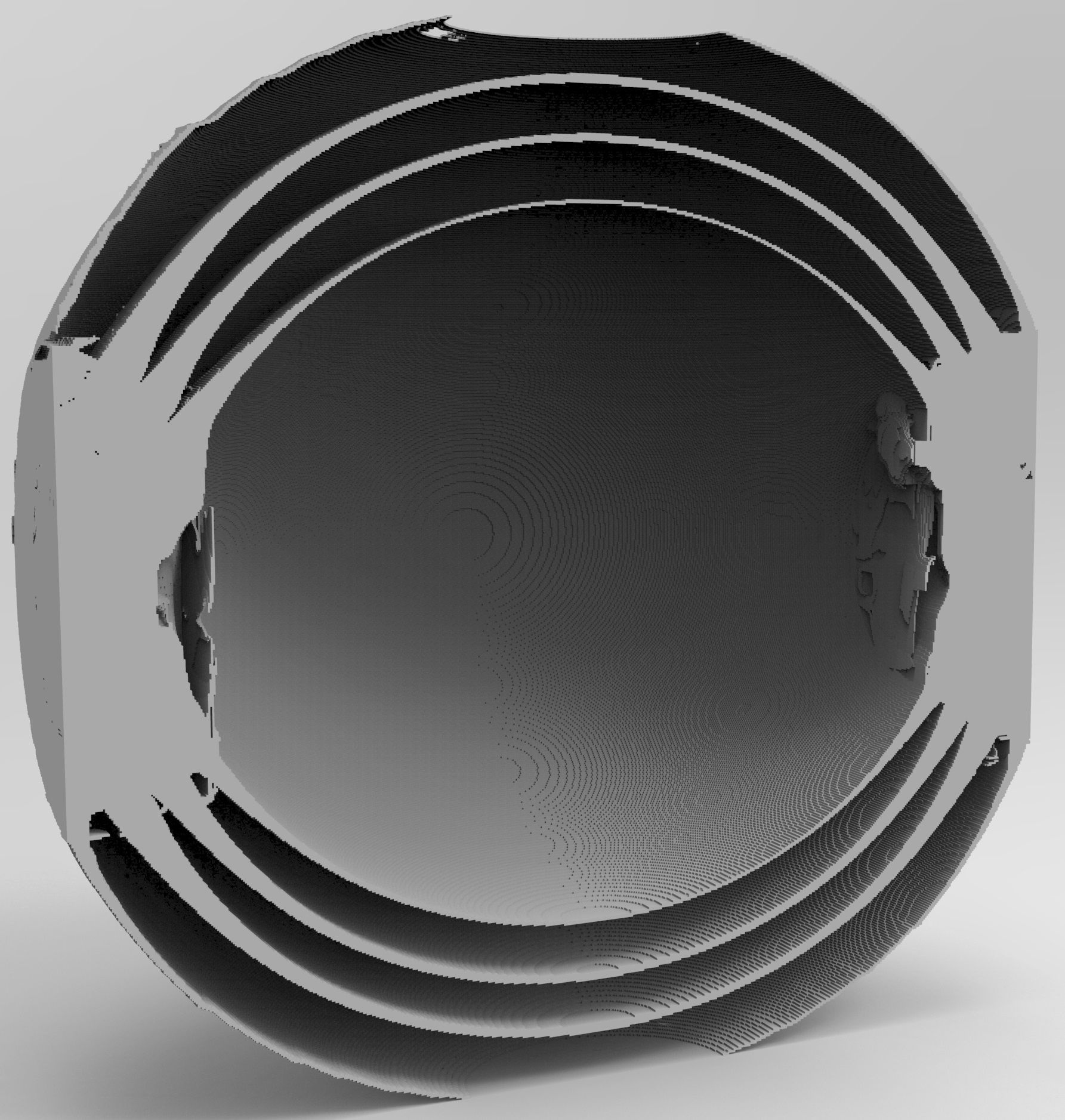}} \hfill \hfill
\caption{Michell's torsion sphere optimized on $\mathcal{T}^{c} = 48\times48\times48$ and de-homogenized on a fine mesh of $576\times576\times576$ using $\varepsilon=32~h^{f}$ and $f_{min} = 2~h^{f}$.}
\label{Fig:DeHomo.4}
\end{figure}

\begin{table}[ht!]
\centering
\caption{The volume fraction $V_{f}^{\phi}$, compliance $\mathcal{J}^{\phi}$, stiffness per weight measure $\mathcal{S}^{\phi}$ and total computation time $T^{tot}$ de-homogenized L-shaped beam on a fine mesh of $768\times768\times384$ elements. Furthermore, the volume fraction $V_{f}^{post}$ and compliance $\mathcal{J}^{post}$ and stiffness per weight  $\mathcal{S}^{post}$ after the post-processing step are shown. Results are shown for different $\mathcal{T}^{c}$, $f_{min}$ and $\varepsilon$.}
\label{Tab:DeHomo.5}
\centering
\setlength{\arrayrulewidth}{1pt} 
\begin{tabular}{ccccccc|ccc}
 \hline
  $\mathcal{T}^{c}$ & $\varepsilon$ & $f_{min}$ & $V_{f}^{\phi}$ & $\mathcal{J}^{\phi}$ & $\mathcal{S}^{\phi}$ &$T^{tot}~[hh:mm:ss]$ & $V_{f}^{post}$ & $\mathcal{J}^{post}$& $\mathcal{S}^{post}$\\ \hline
  $48\times48\times24$& $24~h^{f}$ & $0~h^{f}$ & 0.0945 & 764.85 & 72.309 & 02:26:04 & 0.0934 & 764.93 &71.476  \\
  $48\times48\times24$& $24~h^{f}$ & $2~h^{f}$ & 0.1033 & 716.24 &74.013 &  & 0.1004 & 716.32 &71.947\\
  $48\times48\times24$& $24~h^{f}$ & $3~h^{f}$ & 0.1097 & 689.16 &75.587 &  & 0.1039 & 2484.8 &258.22\\
  $48\times48\times24$& $24~h^{f}$ & $4~h^{f}$ & 0.1164 & 663.32&77.220 &  & 0.1093 & 657.80 &71.881\\
  $48\times48\times24$& $32~h^{f}$ & $0~h^{f}$ & 0.0951 & 833.78&79.321 &  & 0.0943 & 834.47 &78.728\\
  $48\times48\times24$& $32~h^{f}$ & $2~h^{f}$ & 0.1050 & 765.99&80.425 &  & 0.1014 & 766.04 &77.666\\
  $48\times48\times24$& $32~h^{f}$ & $3~h^{f}$ & 0.1004 & 795.00&79.812 &  & 0.0939 & 831.00 &78.006\\
  $48\times48\times24$& $32~h^{f}$ & $4~h^{f}$ & 0.1098 & 736.13&80.859 &  & 0.1036 & 741.82 &76.851\\
  $48\times48\times24$& $40~h^{f}$ & $0~h^{f}$ & 0.0991 & 682.37&67.589 &  & 0.0971 & 682.39&66.274 \\
  $48\times48\times24$& $40~h^{f}$ & $2~h^{f}$ & 0.1031 & 669.78&69.083 &  & 0.0997 & 669.87 &66.756\\
  $48\times48\times24$& $40~h^{f}$ & $3~h^{f}$ & 0.1064 & 659.90&70.249 &  & 0.1010 & 4193.0&423.55 \\
  $48\times48\times24$& $40~h^{f}$ & $4~h^{f}$ & 0.1100 & 651.10&71.615 &  & 0.1030 & 643.14 &66.140\\ 
\hline
  $96\times96\times48$ & $24~h^{f}$ & $0~h^{f}$ & 0.1034 & 593.08 &61.337& 14:31:17 & 0.1031 & 593.08 &61.122\\
  $96\times96\times48$ & $24~h^{f}$ & $2~h^{f}$ & 0.1073 & 582.47& 62.479 &  &  0.1066& 582.56 & 62.083\\
  $96\times96\times48$ & $24~h^{f}$ & $3~h^{f}$ & 0.1107 & 574.35&63.599 &  &   0.1087& 4170.3 &453.17\\
  $96\times96\times48$ & $24~h^{f}$ & $4~h^{f}$ & 0.1143 & 564.24&64.486 &  &  0.1117& 559.72 &62.542\\
  $96\times96\times48$ & $32~h^{f}$ & $0~h^{f}$ & 0.1051 & 599.99&63.037 &  & 0.1048 & 600.10 &62.906\\
  $96\times96\times48$ & $32~h^{f}$ & $2~h^{f}$ & 0.1075 & 590.90&63.546 &  & 0.1070& 590.94 &63.212\\
  $96\times96\times48$ & $32~h^{f}$ & $3~h^{f}$ & 0.1100 & 581.51 &63.986 &  & 0.1090  & 578.38 &63.038\\
  $96\times96\times48$ & $32~h^{f}$ & $4~h^{f}$ & 0.1128 & 571.21&64.433 & & 0.1113 & 567.38 &63.179\\
  $96\times96\times48$ & $40~h^{f}$ & $0~h^{f}$ & 0.0985 & 645.21&63.569 &  & 0.0982& 645.25&63.371\\
  $96\times96\times48$ & $40~h^{f}$ & $2~h^{f}$ & 0.1002 & 638.44&63.974 &  & 0.0998 & 638.46 &63.728\\
  $96\times96\times48$ & $40~h^{f}$ & $3~h^{f}$ & 0.1018 & 633.05&64.490 &  & 0.1008& 29504&2974.7\\
  $96\times96\times48$ & $40~h^{f}$ & $4~h^{f}$ & 0.1038 & 625.70&64.924 &  & 0.1024 & 619.13 &63.377\\ 
\hline
 \end{tabular}
\end{table}

\begin{figure}[h!]
\centering
\hfill
\subfloat[Side view.]{\includegraphics[width=0.55\textwidth]{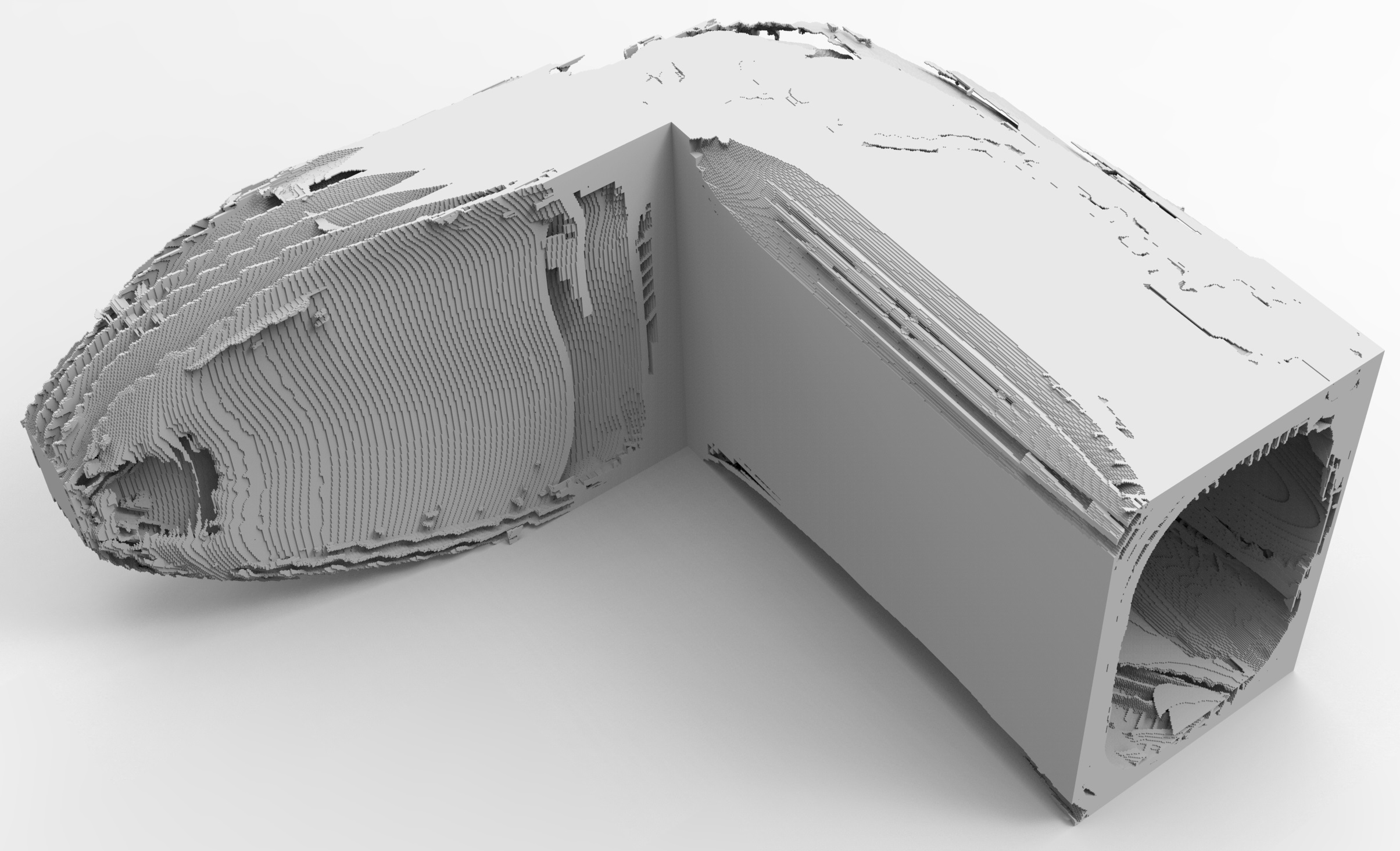}} \hfill
\subfloat[Section view from the top.]{\includegraphics[width=0.30\textwidth]{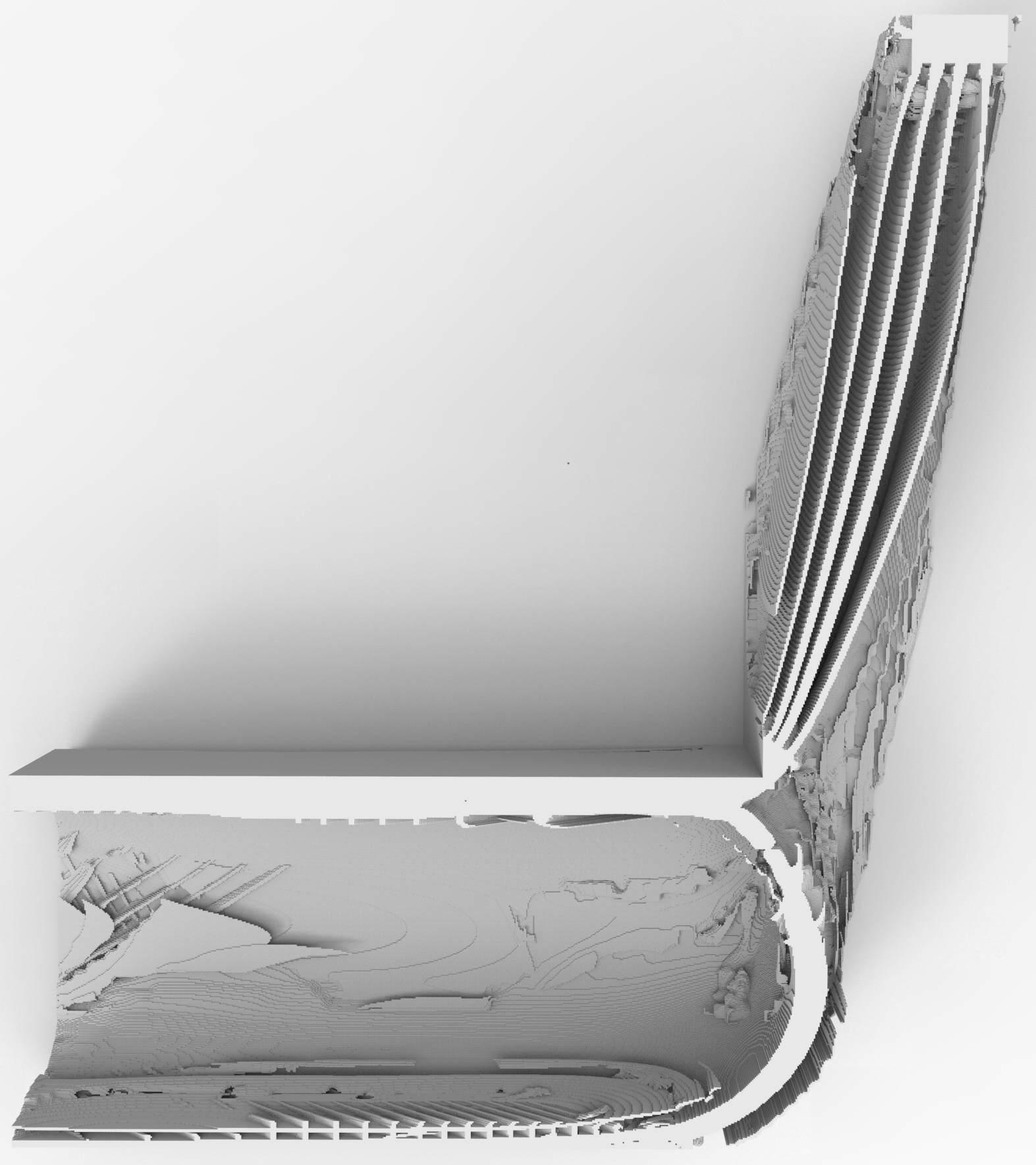}} \hfill \hfill
\caption{The L-shaped beam example optimized on $\mathcal{T}^{c} = 96\times96\times48$ and de-homogenized on a fine mesh of $768\times768\times384$ using $\varepsilon=24~h^{f}$ and $f_{min} = 2~h^{f}$.}
\label{Fig:DeHomo.5}
\end{figure}

As can be seen the Michell cantilever, is very sensitive to the enforcement of a minimum feature size. This is because there are many vertical thin plates just a few voxels wide as can be seen in Figures~\ref{Fig:DeHomo.3}(a),(b) and (c). Hence, the layer widths corresponding to that direction are around $0.1$. This means that for a small unit-cell spacing, the minimum width can be violated and a lot of material is added. Hence, for this example a slightly larger value of $\varepsilon$ is actually better. Furthermore, it can be seen that there is no noticeable difference between the two mesh sizes on which the homogenization-based designs are obtained. Finally, it can be seen that for one case the post-processing scheme resulted in a significantly worse compliance, which is caused by the open-close filter step in which also load carrying material has been removed.

Moving on to Michell's torsion sphere example, we observe a large difference in compliance between the homogenization-based designs and the de-homogenized designs. This is mainly caused by the fact that the load is applied along a line for simplicity, which results in a different loading condition depending on the mesh size. Nevertheless, the goal of this example was to show that the optimized solution visually represents the analytical solution of Michell's torsion sphere which is a close-walled sphere~\citep{Bib:SigmundMichell}. 
Interestingly, this is the case; however, the de-homogenized solution contains several closed spheres of different radii as can be seen in Figure~\ref{Fig:DeHomo.4}. The reason for the multiple spheres is the relatively high volume fraction, the rank-3 material model and the fact that a relatively coarse mesh is used for the homogenization-based topology optimization. Finally, it should be mentioned that using the post-processing procedure a large number of passive solid elements at the boundary condition have been removed, making the comparison with the homogenization-based design unfair.

The orientation field for the L-shaped beam actually contains a singularity at the corner with the passive void domain. However, since this singularity is at the boundary of the domain, the sorting algorithm managed to return smooth and continuous vector fields, although we have to note that this cannot be guaranteed for other cases including singularities. In this example the microstructures and layer orientations are rapidly changing and hence the design obtained on a very coarse optimization mesh resulted in worse performing de-homogenized designs than the ones obtained on $\mathcal{T}^{c} = 96\times96\times48$. As can be seen in Figure~\ref{Fig:DeHomo.5}(a) and (b), the optimized solution is a combination of a cantilever optimized for bending stiffness and a hollow box optimized for torsion. 

\subsection{Comparison with large-scale topology optimization}
Besides verifying that the de-homogenized designs perform similar to the homogenization-based designs, it is interesting to compare the results with a well-established density-based topology optimization method. To do so, we make use of the publicly available topology optimization code using the PETSc framework~\citep{Bib:TopOptPETSc}. The code has been slightly modified to allow for different boundary conditions, passive elements and a continuation scheme on the penalization factor $(p)$. We start with $p=1$ and slowly increase this to $p=4$, and stop the optimization procedure after 450 iterations when the change in design is negligible. Furthermore, it should be noted that the density filter is used in these examples with a filter radius of $R = 3h^{f}$, where the filter is turned off in the final iterations to allow for black and white designs. 

All four examples shown in Figure~\ref{Fig:TOEx.1} are optimized on the same resolution as the de-homogenized designs discussed above. The corresponding fine-scale compliance $\mathcal{J}^{f}$, measure of stiffness per weight $\mathcal{S}^{f}$, volume fraction $V_{f}^{f}$, the number of iterations $n_{iter}$, and run time $T^{f}$ are shown in Table~\ref{Tab:FineSIMP.1}. Furthermore, the design for the cantilever, and the electrical mast example are shown in Figures~\ref{Fig:DeHomo.2}(c) and (d) respectively. Finally it is mentioned that the same high-performance computing cluster is used as for the fine scale validations. Hence, for each optimization example 100 nodes each with 32 cores are used.

\begin{table}[ht!]
\centering
\caption{Compliance $\mathcal{J}_{f}$, volume fraction $V^{f}_{f}$, the number of iterations $n_{iter}$, and run time $T_{f}$ for different optimization examples optimized using a density-based optimization procedure.}
\label{Tab:FineSIMP.1}
\centering
\setlength{\arrayrulewidth}{1pt} 
\begin{tabular}{ccccccc}
 \hline
 \textbf{Example} & \textbf{Mesh size} & $\mathcal{J}^{f}$ & $\mathcal{S}^{f}$ &  $V_{f}^{f}$ & $n_{iter}$ & $T^{f}$ \\ \hline
 Michell cantilever & $960\times480\times480$ & $236.262$& $23.626$ & $0.100$  & 450& 08:13:28 \\
 Michell's torsion sphere & $576\times576\times576$ & $20.916$& $2.0916$ & $0.100$  & 450& 01:03:06 \\
 Electrical mast & $384\times384\times1152$ & $99.961$ & $9.9961$ & $0.100$  & 450& 02:16:28  \\
 L-beam  & $768\times768\times384$ & $585.114$ & $58.511$& $0.100$  & 450& 04:38:50 \\
\hline
 \end{tabular}
\end{table}

It can be seen that the compliance values of the density-based designs are close in general even $5-10\%$ better than the de-homogenized designs. This is more or less similar when we compare the stiffness per weight measure. The main reason is that the de-homogenized designs represents a multi-scale structure, which ideally should be de-homogenized on a finer mesh, to accurately capture all these multi-scale features. But more importantly, the well-performing de-homogenized designs can be obtained using a modern PC, while for the density-based designs an expensive high-performance computing cluster is required. The de-homogenization procedure can therefore reduce the threshold for using topology optimization for generating large-scale designs. By looking at the run time and number of cores we can conclude that the total computational cost can be reduced by at least 3 orders of magnitude compared to standard density-based topology optimization!
\section{Concluding remarks}
\label{Sec:Conclusion}
We have presented a highly efficient approach to obtain manufacturable and ultra-high resolution designs from homogenization-based topology optimization. By doing homogenization-based topology optimization using optimal microstructures, an optimal design can be represented on a relatively coarse mesh. A good rule of thumb would be that at least 48 elements in each direction of the mesh are required. The subsequently de-homogenized designs perform within $5-10\%$ compared to designs obtained using well-established density-based topology optimization. However, instead of using high performance computing facilities with more than 3000 cores the presented designs have been obtained using a single core MATLAB process on a modern workstation PC. Hence, the presented procedure is a first step on the way to achieve giga-scale interactive topology optimization.

Besides obtaining near-optimal designs, we have presented a method to control the shape and minimum feature size of these de-homogenized designs ensuring manufacturability. From the results it can also be seen that even better performing designs can be obtained when de-homogenizing on a finer mesh. This paves the way for future studies into different methods to represent the de-homogenized designs, other than the voxel grid that has been used now. Note however that the presented procedure only works for examples where the homogenization-based designs are free of singularities. Hence, a natural extension would be to better understand the occurence of singularities and extend the de-homogenization procedure such that it takes these singularities into account. We are confident that this can be done to make topology optimization an integrated part of the design process for large-scale problems.

\section*{Acknowledgments}
The authors acknowledge the financial support from the Villum Foundation (InnoTop VILLUM investigator project). Furthermore, the authors would like to express their gratitude to the members of the TopOpt group at DTU for valuable discussions during the preparation of this work. Finally, the authors wish to thank Krister Svanberg for the Matlab MMA code.
\bibliography{DataBase}   
\bibliographystyle{elsarticle-harv}\biboptions{authoryear}

\end{document}